\documentclass[twocolumn,showpacs,showkeys]{revtex4}
\usepackage{graphicx}
\usepackage{bm}
\usepackage{color}
\usepackage{amsmath}
\usepackage{natbib}

\begin{document}

\title{Spin-electron acoustic waves: The Landau damping and ion contribution in the spectrum}

\author{Pavel A. Andreev}
\email{andreevpa@physics.msu.ru}
\affiliation{Faculty of physics, Lomonosov Moscow State University, Moscow, Russian Federation.}

 \date{\today}

\begin{abstract}
Separated spin-up and spin-down quantum kinetics is derived for more detailed research of the spin-electron acoustic waves. Kinetic theory allows to obtain spectrum of the spin-electron acoustic waves including effects of occupation of quantum states more accurately than quantum hydrodynamics. We apply quantum kinetic to calculate the Landau damping of the spin-electron acoustic waves. We have considered contribution of ions dynamics in the spin-electron acoustic wave spectrum. We obtain contribution of ions in the Landau damping in temperature regime of classic ions. Kinetic analysis for ion-acoustic, zero sound, and Langmuir waves at separated spin-up and spin-down electron dynamics is presented as well.
\end{abstract}

\pacs{52.30.Ex, 52.35.Dm}% PACS, the Physics and Astronomy
                             % Classification Scheme.
\keywords{quantum plasmas, quantum kinetics, wave dispersion, Landau damping, spin-electron acoustic wave}
%Use showkeys class option if keyword

\maketitle

%52.30.Ex	Two-fluid and multi-fluid plasmas
%52.35.Dm	Sound waves

%52.27.Ep	Electron-positron plasmas

%%%%%%%%%%TEXT

\section{\label{sec:level1} Introduction}

Recently developed separate spin evolution quantum hydrodynamic (SSE-QHD) model \cite{Andreev spin-up and spin-down 1405}, giving separated description of spin-up and spin-down electrons, allowed us to discover new type of longitudinal collective excitations in degenerate quantum plasmas. This excitation is called the spin-electron acoustic wave (SEAW). In this model, spin-up and spin-down electrons are considered as two different species. The SEAW exists in magnetised plasmas due to difference of the Fermi momentum of spin-up and spin-down electrons at presence of an external magnetic field.

Propagation of the SEAWs parallel and perpendicular to an external magnetic field was considered in Ref. \cite{Andreev spin-up and spin-down 1405}. Further research of the SEAWs was performed in Refs. \cite{Andreev spin-up and spin-down 1406 Oblique} and \cite{Andreev spin-up and spin-down 1408 2D}. Dispersion of the SEAWs in different two dimensional structures was studied in Ref. \cite{Andreev spin-up and spin-down 1408 2D}. Plane-like two dimensional electron gas in a magnetic field perpendicular to the sample, and conducting nanotubes, having cylindrical geomentry, in an external magnetic field parallel to the cylinder axis were considered in Ref. \cite{Andreev spin-up and spin-down 1408 2D}.

More general case of oblique wave propagation in three dimensional structures was considered in Ref. \cite{Andreev spin-up and spin-down 1406 Oblique}. It was demonstrated that at oblique propagation we have two branches of the SEAWs instead of one existing in limit cases of parallel and perpendicular propagation.

In paper \cite{Andreev spin-up and spin-down 1405}, derivation of the separated spin evolution QHD with two different species of spin-up and spin-down electrons was demonstrated on a simple example of the single-particle Pauli equation. It rather obvious that a single-particle equation has nothing to do with a plasma description. A full derivation should be based upon a many-particle theory, as, for instance, the many-particle QHD (MPQHD) developed by Kuz'menkov and coauthors \cite{MaksimovTMP 2001}, \cite{MaksimovTMP 2001 b}, \cite{Andreev RPJ 07}, \cite{Andreev AtPhys 08}, \cite{Andreev PRB 11}, \cite{Andreev spin current}. The final equations, presented in Ref. \cite{Andreev spin-up and spin-down 1405}, were obtained by the corresponding modification of the MPQHD. Hence they have more general form than the result of the separate spin-up and spin-down fluidisation of the single-particle Pauli equation. Nevertheless application of the single-particle Pauli equation was a simple way to demonstrate the correct structure of the SSE-QHD. This is a generalisation of famous fluidisation of the single-particle Pauli equation performed by Takabayasi \cite{Takabayasi PTP 55 b}.

Consequences of separate spin evolution for the Langmuir \cite{Andreev spin-up and spin-down 1405}, \cite{Andreev spin-up and spin-down 1406 Oblique}, \cite{Andreev spin-up and spin-down 1408 2D} and Trivelpiece--Gould \cite{Andreev spin-up and spin-down 1406 Oblique} waves were also studied in mentioned parers.

This paper is dedicated to further analysis of the spin-electron acoustic waves and influence of separate spin evolution of electrons on ion acoustic and zeroth sound waves.
In this paper we focus our attention on waves propagating parallel to the external magnetic field.
Here we develop the separate spin evolution quantum kinetics. Kinetic theory gives a background for more careful analysis of distribution of electrons over different quantum states and contribution of these effects in the quantum plasma properties.

Since we have an example of derivation of SSE-QHD from the single particle Pauli equation, we stress attention on a many-particle derivation of separate spin evolution quantum kinetics.

The separate spin evolution quantum kinetics allows to calculate the Landau damping of the SEAWs. We apply the separate spin evolution quantum kinetics to obtain the Landau damping of the SEAWs and other excitations in two different regimes: regime of intermediate temperatures, when electrons are degenerate and ions are classical, and regime of low temperatures when all species are degenerate.

Some topics in quantum plasmas were discussed in reviews \cite{Uzdensky arxiv review 14}, \cite{Shukla RMP 11}, \cite{Andreev 1407 Review}.

This paper is organized as follows.  In Sec. II we describe basic definitions of quantum kinetics and describe quantum mechanic background essential for derivation of the quantum kinetics.  Sec. III contains closed set of separate spin evolution quantum kinetic equations. In Sec. IV equilibrium state is described. Linearised kinetic equations for small perturbations of the equilibrium are presented in Sec. IV as well. In Sec. V a general form of dispersion equation for oblique propagating longitudinal waves is obtained. In Sec. VI we present detailed analysis of spectrum of longitudinal waves propagating parallel to the external magnetic field. In Sec. VII a brief summary of obtained results is presented.

\section{\label{sec:level1} Method of derivation of Separated spin-up and spin-down quantum kinetics}

\subsection{Structure of many-particle N-spinor wave function}

If we have a single particle with no spin degree of freedom it can be described by the wave function $\psi(\textbf{r},t)$, which is a complex function of three space coordinates and time. If we have two particles of that kind we need to apply the two-particle wave function $\psi(\textbf{r}_{1}, \textbf{r}_{2},t)$, which is a complex function in six dimensional configuration space. It is hard to make assumptions for this function when we consider two interacting particles. However if interaction is weak, or we have two non-interacting particles, we can represent a two-particle wave function as the product of single particle wave functions $\psi(\textbf{r}_{1}, \textbf{r}_{2},t)=\psi(\textbf{r}_{1},t) \psi(\textbf{r}_{2},t)$, or including antisymmetry of fermion wave function $\psi(\textbf{r}_{1}, \textbf{r}_{2},t)=\frac{1}{2}\left|
                                                                                                     \begin{array}{cc}
                                                                                                       \psi_{a_{1}}(\textbf{r}_{1},t) & \psi_{a_{1}}(\textbf{r}_{2},t) \\
                                                                                                       \psi_{a_{2}}(\textbf{r}_{1},t) & \psi_{a_{2}}(\textbf{r}_{2},t) \\
                                                                                                     \end{array}
                                                                                                   \right|
$.

Next focus our attention on the spin-1/2 particles, as electrons, which are main subject of this paper. A single spin-1/2 particle is descried by the spinor (the first-rank spinor) wave function, which is a two component wave function $\psi=\left(
                                                                                             \begin{array}{c}
                                                                                               \psi_{u} \\
                                                                                               \psi_{d} \\
                                                                                             \end{array}
                                                                                           \right)
$ (see for instance \cite{Landau Vol 3} section 56). Next step on the way of construction of the many-particle theory is obtaining of the wave function for two spin-1/2 particles. This wave function appears to be a second rank spinor \cite{Landau Vol 3}.

Second-rank spinor is a four component quantity $\psi_{\varsigma\tau}$ (see for instance \cite{Landau Vol 3} section 56 after formula 56.13). Components of $\psi_{\varsigma\tau}$ are transformed as products $\psi_{\varsigma}\psi_{\tau}$ of components of two first-rank spinors. The $2\times2$ unit matrix $\hat{I}$ together with three sigma (Pauli) matrixes form a basis in space of the second-rank spinors.

Summarising all written above we can present a scheme of generalisation:
\begin{equation}\label{SUSDKin WF generalisation}\left(
                                 \begin{array}{ccc}
                                   \psi(\textbf{r},t) & \Rightarrow & \psi(\textbf{r}_{1}, \textbf{r}_{2},t) \\
                                   \Downarrow &   & \Downarrow \\
                                   \left(
                                     \begin{array}{c}
                                       \psi_{u}(\textbf{r},t) \\
                                       \psi_{d}(\textbf{r},t) \\
                                     \end{array}
                                   \right)
                                    & \Rightarrow & \Psi(\textbf{r}_{1}, \textbf{r}_{2},t), \\
                                 \end{array}
                               \right)
\end{equation}
where
\begin{equation}\label{SUSDKin two part WF of spin part} \Psi(\textbf{r}_{1}, \textbf{r}_{2},t)=\left(
                                                                                 \begin{array}{cc}
                                                                                   \psi_{u1}(\textbf{r}_{1}, \textbf{r}_{2},t) & \psi_{u2}(\textbf{r}_{1}, \textbf{r}_{2},t) \\
                                                                                   \psi_{d1}(\textbf{r}_{1}, \textbf{r}_{2},t) & \psi_{d2}(\textbf{r}_{1}, \textbf{r}_{2},t), \\
                                                                                 \end{array}
                                                                               \right)
\end{equation}
is the second rank spinor having presentation of $2\times2$ matrix. The two-particle wave function of spin-1/2 particles, being a matrix, should wear spinor subindexes, hence we write $\Psi(\textbf{r}_{1}, \textbf{r}_{2},t)=\Psi_{s_{1}s_{2}}(\textbf{r}_{1}, \textbf{r}_{2},t)$.

Spin and coordinate parts of the many-particle wave function can be separated in absence of the spin-current and spin-orbit interactions
\begin{equation}\label{SUSDKin} \Psi(R,t)=\Psi_{S}(R,t)=\Lambda(R,t) \cdot \chi_{S}, \end{equation}
where $R=\{\textbf{r}_{1}, ..., \textbf{r}_{i}, ..., \textbf{r}_{N}\}$ is the set of coordinates of $N$ particles, $S=\{s_{1}, ..., s_{i}, ..., s_{N}\}$ is the set of spinor subindexes of $N$ particles.
Full many-particle wave function $\Psi(R,t)$ is antisymmetric to permutation of arguments. Hence if orbital part $\Lambda(R,t)$ is antisymmetric, then the spin part $\chi_{S}$ is symmetric. In opposite case orbital part $\Lambda(R,t)$ is symmetric, then the spin part $\chi_{S}$ is antisymmetric.

\subsection{Many-particle Pauli equation as the starting point of derivation of kinetic equations}

Thus we start with the many-particle Pauli equation
$$ \imath\hbar\partial_{t}\Psi(R,t)=\biggl(\sum_{i=1}^{N}\biggl(\frac{1}{2m_{i}}\hat{\textbf{D}}_{i}^{2} +q_{i}\varphi^{ext}_{i}-\gamma_{i}\mbox{\boldmath $\sigma$}_{i}\textbf{B}_{i(ext)}\biggr)$$
\begin{equation}\label{SUSDKin Pauli} +\frac{1}{2}\sum_{i,j\neq i}^{N}(q_{i}q_{j}G_{ij}-\gamma_{i}\gamma_{j}G^{\alpha\beta}_{ij}\sigma^{\alpha}_{i}\sigma^{\beta}_{j})\biggr)\Psi(R,t), \end{equation}
where $\Psi(R,t)=\Psi_{S}(R,t)$, we can also rewrite terms containing spin operators with detail description of spinor indexes $\mbox{\boldmath $\sigma$}_{i}\Psi(R,t)=(\mbox{\boldmath $\sigma$}_{i}\Psi(R,t))_{S}=(\mbox{\boldmath $\sigma$}_{i}\Psi(R,t))_{s_{1}, ..., s_{i}, ..., s_{N}}=\mbox{\boldmath $\sigma$}_{s_{i}s_{i'}}\Psi_{s_{1}, ..., s_{i'}, ..., s_{N}}(R,t)$,  and $\sigma^{\alpha}_{i}\sigma^{\beta}_{j}\Psi(R,t)=(\sigma^{\alpha}_{i}\sigma^{\beta}_{j}\Psi(R,t))_{S} =(\sigma^{\alpha}_{i}\sigma^{\beta}_{j}\Psi(R,t))_{s_{1}, ..., s_{i}, ..., s_{j}, ..., s_{N}} =\sigma^{\alpha}_{s_{i}s_{i'}}\sigma^{\beta}_{s_{j}s_{j'}}\Psi_{s_{1}, ..., s_{i'}, ..., s_{j'}, ..., s_{N}}(R,t)$.

Equation (\ref{SUSDKin Pauli}) governs evolution of N-spinor wave function $\Psi(R,t)$. In equation (\ref{SUSDKin Pauli}) $m_{i}$ is the mass of particle with number $i$, below we consider system of particles with equal masses, $q_{i}$ is the charge of particle, $\gamma_{i}$ is the gyromagnetic ratio, for electrons it can be written as $\gamma_{i}\approx1.00116\mu_{B}$, $\mu_{B}=q_{i}\hbar/(2m_{i}c)$ is the Bohr magneton, the difference of $\mid\gamma_{e}\mid$ from the Bohr magneton includes contribution of the anomalous magnetic dipole moment, $\varphi^{ext}_{i}$ is the scalar potential of an external electromagnetic field acting on particle with number $i$, $\textbf{B}_{i(ext)}$ is the external magnetic field,
$(\hat{\textbf{D}}_{i}\psi)(R,t)=((-\imath\hbar\nabla_{i}-\frac{q_{i}}{c}\textbf{A}_{i,ext})\psi)(R,t)$, with $\textbf{A}_{i,ext}$ is the vector potential of an external electromagnetic field acting on particle, $\mbox{\boldmath $\sigma$}_{i}$ is the Pauli matrixes describing spin of particles, $\hbar$ is the reduced Planck constant, $c$ is the speed of light,
$\textbf{B}_{i(ext)}=\nabla_{i}\times \textbf{A}_{i(ext)}$,
$\textbf{E}_{i(ext)}=-\nabla_{i}\varphi_{ext}(\textbf{r}_{i},t)-\frac{1}{c}\partial_{t}\textbf{A}_{ext}(\textbf{r}_{i},t)$, $G_{pn}=\frac{1}{r_{ij}}$ is the Green function of the Coulomb interaction containing module of the interparticle distance $\textbf{r}_{ij}=\textbf{r}_{i}-\textbf{r}_{j}$, and
\begin{equation}\label{SUSDKin} G^{\alpha\beta}_{ij}=4\pi\delta_{\alpha\beta}\delta(\textbf{r}_{ij})+\nabla^{\alpha}_{i}\nabla^{\beta}_{i}\frac{1}{r_{ij}}\end{equation}
is the Green function of the spin-spin interaction, $\delta_{ij}$ is the Kroneckers delta.

Let us present the explicit form of the Pauli matrixes
\begin{equation}\label{SUSDKin Pauli Matrixes}\begin{array}{ccc} \widehat{\sigma}_{x}=\left(\begin{array}{ccc}0& 1\\
1& 0\\
\end{array}\right),&
\widehat{\sigma}_{y}=\left(\begin{array}{ccc}0& -\imath \\
\imath & 0 \\
\end{array}\right),&
\widehat{\sigma}_{z}=\left(\begin{array}{ccc} 1& 0\\
0& -1\\
\end{array}\right).
\end{array}\end{equation}
The commutation relation for spin-1/2 matrixes is
\begin{equation}\label{SUSDKin comm rel} [\widehat{\sigma}^{\alpha}, \widehat{\sigma}^{\beta}]=2\imath\varepsilon^{\alpha\beta\gamma}\widehat{\sigma}^{\gamma},\end{equation}
with $\varepsilon^{\alpha\beta\gamma}$ is the Levi-Civita symbol.

\subsection{Explicit form of the Pauli equation for two spin-1/2 particles}

As the first simple example let us rewrite the Pauli equation (\ref{SUSDKin Pauli}) for a single particle in more explicit form
$$\imath\hbar\partial_{t}\psi_{\uparrow}=\biggl(\frac{(\frac{\hbar}{\imath}\nabla-\frac{q_{e}}{c}\textbf{A})^{2}}{2m}+q_{e}\varphi-\gamma_{e}B_{z}\biggr)\psi_{\uparrow}$$
\begin{equation}\label{SUSDKin Pauli Expl +}
-\gamma_{e}(B_{x}-\imath B_{y})\psi_{\downarrow}, \end{equation}
and
$$\imath\hbar\partial_{t}\psi_{\downarrow}=\biggl(\frac{(\frac{\hbar}{\imath}\nabla-\frac{q_{e}}{c}\textbf{A})^{2}}{2m}+q_{e}\varphi+\gamma_{e}B_{z}\biggr)\psi_{\downarrow}$$
\begin{equation}\label{SUSDKin Pauli Expl -}
-\gamma_{e}(B_{x}+\imath B_{y})\psi_{\uparrow}. \end{equation}

The single particle wave spinor can be easily presented as a linear combination of the spin-up and spin-down states described corresponding unit spinors
\begin{equation}\label{SUSDKin Spinors One part} \psi_{s}(\textbf{r},t)=\psi_{\uparrow}\left(\begin{array}{ccc}1\\
0\\
\end{array}\right)
+\psi_{\downarrow}\left(\begin{array}{ccc}0 \\
1 \\
\end{array}\right).
\end{equation}
It have allowed us to rewrite the Pauli equation in a rather more explicit form given be formulae (\ref{SUSDKin Pauli Expl +}) and (\ref{SUSDKin Pauli Expl -}).

Applying wave functions describing spin-up $\psi_{\uparrow}$ and spin-down $\psi_{\downarrow}$ states we can write probability density to find the particle in a point $\textbf{r}$ with spin-up $\rho_{\uparrow}=\mid\psi_{\uparrow}\mid^{2}$ or spin-down $\rho_{\downarrow}=\mid\psi_{\downarrow}\mid^{2}$. We also see $\rho=\rho_{\uparrow}+\rho_{\downarrow}$. Directions up $\uparrow$ (down $\downarrow$) corresponds to spins having same (opposite) direction as (to) the external magnetic field. While magnetic moments have opposite to spin directions since we consider negatively charged particles.

Let us consider structure of the many-particle wave function $\Psi(R,t)$ in more details. To our end we will need different basis in space of the second-rank spinors:

\begin{equation}\label{SUSDKin Spinors Two part 1  1}
\tau_{1}=\frac{1}{2} \biggl(\widehat{\sigma}_{0}+\widehat{\sigma}_{z}\biggr)=\left(\begin{array}{ccc} 1& 0\\
0& 0\\
\end{array}\right)\end{equation}

\begin{equation}\label{SUSDKin Spinors Two part 1  2}\tau_{2}= \frac{1}{2} \biggl(\widehat{\sigma}_{x}+\imath\widehat{\sigma}_{y}\biggr)=\left(\begin{array}{ccc}0& 1\\
0& 0\\
\end{array}\right)\end{equation}

\begin{equation}\label{SUSDKin Spinors Two part 1  3}
\tau_{3}=\frac{1}{2} \biggl(\widehat{\sigma}_{x}-\imath\widehat{\sigma}_{y}\biggr)=\left(\begin{array}{ccc}0& 0 \\
1 & 0 \\
\end{array}\right),\end{equation}
and
\begin{equation}\label{SUSDKin Spinors Two part 1  4}
\tau_{4}=\frac{1}{2} \biggl(\widehat{\sigma}_{0}-\widehat{\sigma}_{z}\biggr)=\left(\begin{array}{ccc} 0& 0\\
0& 1\\
\end{array}\right),\end{equation}
where
\begin{equation}\label{SUSDKin Spinors Two part 2}\widehat{\sigma}_{0}=\hat{I}=\left(\begin{array}{ccc}1& 0\\
0& 1\\
\end{array}\right) \end{equation}
is the unit second rank spinor.

It is essential to repeat that the two-particle wave function of two spin-1/2 particles is a $2\times2$ matrix (see formula (\ref{SUSDKin two part WF of spin part})). Consequently we can expand it as a superposition of matrixes $\widehat{\sigma}_{0}$ and $\mbox{\boldmath $\sigma$}=\{\widehat{\sigma}_{x}, \widehat{\sigma}_{y}, \widehat{\sigma}_{z}\}$ or the set of matrixes $\{ \widehat{\tau}_{a} \}$ with $a=1$, $2$, $3$, $4$. So, the two-particle wave function has form of $\Psi =\sum_{a}\psi_{a} \widehat{\tau}_{a}$, where $\psi_{a}=\psi_{a}(\textbf{r}_{1}, \textbf{r}_{2}, t)$ are complex functions.

Wave function of two spin-1/2 particles $\Psi_{S}(R,t)$ can be presented via the upper $\Psi_{\uparrow}(R,t)$, or lower $\Psi_{\downarrow}(R,t)$ line in the 2-rank spinor $\Psi_{S}(\textbf{r}_{1}, \textbf{r}_{2},t)=\left(
                                                                                                     \begin{array}{c}
                                                                                                       \Psi_{\uparrow}(\textbf{r}_{1}, \textbf{r}_{2},t) \\
                                                                                                       \Psi_{\downarrow}(\textbf{r}_{1}, \textbf{r}_{2},t) \\
                                                                                                     \end{array}
                                                                                                   \right)$, where $\Psi_{\uparrow}(\textbf{r}_{1}, \textbf{r}_{2},t)=\left(
                                                                                                                       \begin{array}{cc}
                                                                                                                         \psi_{1}(\textbf{r}_{1}, \textbf{r}_{2},t) & \psi_{2}(\textbf{r}_{1}, \textbf{r}_{2},t) \\
                                                                                                                       \end{array}
                                                                                                                     \right)
                                                                                                   $, and $\Psi_{\downarrow}(\textbf{r}_{1}, \textbf{r}_{2},t) =\left(
                                                                                                              \begin{array}{cc}
                                                                                                                \psi_{3}(\textbf{r}_{1}, \textbf{r}_{2},t) & \psi_{4}(\textbf{r}_{1}, \textbf{r}_{2},t) \\
                                                                                                              \end{array}
                                                                                                            \right)
                                                                                                   $.

The density probability in the six dimensional configurational space appears in the traditional form $\rho(\textbf{r}_{1}, \textbf{r}_{2},t)=\Psi^{+}(R,t)\Psi(R,t)$. Its explicit form is $\rho(\textbf{r}_{1}, \textbf{r}_{2},t)=\sum_{i}\psi_{i}(\textbf{r}_{1}, \textbf{r}_{2},t)$. We can separate terms in this sum in two groups $\rho_{\uparrow}(\textbf{r}_{1}, \textbf{r}_{2},t)$ and $\rho_{\downarrow}(\textbf{r}_{1}, \textbf{r}_{2},t)$ in the following way. We can introduce a notation $\rho_{\uparrow}(\textbf{r}_{1}, \textbf{r}_{2},t)=\Psi_{\uparrow}^{+}\Psi_{\uparrow}$, $\rho_{\downarrow}(\textbf{r}_{1}, \textbf{r}_{2},t)=\Psi_{\downarrow}^{+}\Psi_{\downarrow}$, which is not a mathematical symbol, but it will be very useful to get a compact form of formulae below.
Applying wave functions describing spin-up $\Psi_{\uparrow}$ and spin-down $\Psi_{\downarrow}$ states of two particle systems we can write probability density to find both particles in point $\textbf{r}_{1}$ and $\textbf{r}_{2}$ with spin-up $\rho_{\uparrow}(\textbf{r}_{1}, \textbf{r}_{2},t)=\mid\psi_{1}\mid^{2}+\mid\psi_{2}\mid^{2}$ or spin-down $\rho_{\downarrow}=\mid\psi_{3}\mid^{2}+\mid\psi_{4}\mid^{2}$. We also see $\rho=\rho_{\uparrow}+\rho_{\downarrow}$. Directions up $\uparrow$ (down $\downarrow$) corresponds to spins having same (opposite) direction as (to) the external magnetic field.

A compact form of the Pauli equation for two spin-1/2 interacting particles can be written easily using the general form of the Pauli equation for N particles (\ref{SUSDKin Pauli})
$$ \imath\hbar\partial_{t}\Psi(R_{2},t)=\biggl[\sum_{i=1}^{2}\biggl(\frac{1}{2m_{i}}\hat{\textbf{D}}_{i}^{2} +q_{i}\varphi^{ext}_{i}-\gamma_{i}\mbox{\boldmath $\sigma$}_{i}\textbf{B}_{i(ext)}\biggr)$$
\begin{equation}\label{SUSDKin Pauli 2 particle short form} +q_{1}q_{2}G_{12} -\gamma_{1}\gamma_{2}G^{\alpha\beta}_{12}\sigma^{\alpha}_{1}\sigma^{\beta}_{2})\biggr]\Psi(R,t), \end{equation}
where $\Psi(R_{2},t)=\Psi_{s_{1},s_{2}}(\textbf{r}_{1},\textbf{r}_{2},t)$.

We are going to present a more explicit form of equation (\ref{SUSDKin Pauli 2 particle short form}). To this end we introduce operator $\widehat{\Lambda}$ as
\begin{equation}\label{SUSDKin Lambda} \widehat{\Lambda}=\imath\hbar\partial_{t}-\sum_{i=1}^{2}\biggl(\frac{1}{2m_{i}}\hat{\textbf{D}}_{i}^{2} +q_{i}\varphi^{ext}_{i}\biggr)-q_{1}q_{2}G_{12}. \end{equation}

Finally we able to present explicit form of the Pauli equation for two particles involved in the Coulomb and spin-spin interactions, and also interacting with the external electromagnetic field
\begin{widetext}
$$\widehat{\tau}_{1}\biggl\{\Lambda\psi_{1}+\gamma_{1}[B_{1z}\psi_{1}+(B_{1x}-\imath B_{1y})\psi_{3}] +\gamma_{1}\gamma_{2}[(G_{zx}+G_{zy}+G_{zz})\psi_{1}+(G_{xx}+G_{xy}+G_{xz})\psi_{3}-\imath(G_{yx}+G_{yy}+G_{yz})\psi_{3}]\biggr\}$$
$$+\widehat{\tau}_{2}\biggl\{\Lambda\psi_{2}+\gamma_{2}[B_{2z}\psi_{2}+(B_{2x}-\imath B_{2y})\psi_{4}] +\gamma_{1}\gamma_{2}[(G_{xz}+G_{yz}+G_{zz})\psi_{2}+(G_{xx}+G_{yx}+G_{zx})\psi_{4}-\imath(G_{xy}+G_{yy}+G_{zy})\psi_{4}]\biggr\}$$
$$+\widehat{\tau}_{3}\biggl\{\Lambda\psi_{3}+\gamma_{1}[-B_{1z}\psi_{3}+(B_{1x}+\imath B_{1y})\psi_{1}] +\gamma_{1}\gamma_{2}[-(G_{zx}+G_{zy}+G_{zz})\psi_{3}+(G_{xx}+G_{xy}+G_{xz})\psi_{1}+\imath(G_{yx}+G_{yy}+G_{yz})\psi_{1}]\biggr\}$$
\begin{equation}\label{SUSDKin Pauli 2 particle Explicit form}     +\widehat{\tau}_{4}\biggl\{\Lambda\psi_{4}+\gamma_{2}[-B_{2z}\psi_{4}+(B_{2x}+\imath B_{2y})\psi_{2}] +\gamma_{1}\gamma_{2}[-(G_{xz}+G_{yz}+G_{zz})\psi_{4}+(G_{xx}+G_{yx}+G_{zx})\psi_{2}+\imath(G_{xy}+G_{yy}+G_{zy})\psi_{2}]\biggr\}=0. \end{equation}
\end{widetext}

Comparing formula (\ref{SUSDKin Pauli 2 particle Explicit form}) with its analog for a single particle presented by equations (\ref{SUSDKin Pauli Expl +}) and (\ref{SUSDKin Pauli Expl -}) we see that two particle system is rather more complicate. It is essential to mention that two particle Pauli equation reflects many properties of N particle Pauli equation. Hence it allows to understand correct structure of quantum kinetics of electrons with separated spin-up and spin down evolution.

\subsection{Basic definitions of quantum kinetics}

Most famous quantum distribution function was suggested by Wigner \cite{Wigner PR 84}, however we do not apply it here. We construct our kinetic theory in according with the many-particle quantum hydrodynamic (MPQHD) method \cite{MaksimovTMP 2001}, \cite{Andreev RPJ 07}, \cite{Andreev spin current}. We start with classic microscopic distribution function \cite{Weinberg book}, \cite{Klimontovich book}. We change classic dynamic functions of position and momentum of particles on the corresponding operators. As the result we find the operator of many-particle microscopic quantum distribution function \cite{Andreev arXiv 14 positrons},
\cite{Andreev kinetics 13}
\begin{equation}\label{SUSDKin} \hat{f}=\sum_{i}\delta(\textbf{r}-\widehat{\textbf{r}}_{i})\delta(\textbf{p}-\widehat{\textbf{p}}_{i}).\end{equation}

Quantum mechanical averaging of the operator of many-particle distribution function gives us the microscopic distribution function for system of spinning particles \cite{Andreev arXiv 14 positrons},
\cite{Andreev kinetics 13}
$$f_{a}(\textbf{r}, \textbf{p},t)=\frac{1}{4}\int \Biggl(\Psi^{+}(R,t)\sum_{i}\biggl(\delta(\textbf{r}-\textbf{r}_{i})\delta(\textbf{p}-\widehat{\textbf{p}}_{i})$$
\begin{equation}\label{SUSDKin def distribution function el all} +\delta(\textbf{p}-\widehat{\textbf{p}}_{i})\delta(\textbf{r}-\textbf{r}_{i})\biggr)\Psi(R,t)+h.c.\Biggr)dR,\end{equation}
for each species of particles $a=e$ for electrons and $a=i$ for ions. In formula (\ref{SUSDKin def distribution function el all}) we have $\Psi^{+}(R,t)=\Psi_{S}^{+}(R,t)=(\Psi^{S}(R,t))^{*}$.

We can introduce the distribution function of subspecies of electrons for spin-up and spin-down electrons:
$$f_{e,s}=f_{e,s}(\textbf{r},\textbf{p},t)=\frac{1}{4}\int \Biggl(\Psi^{+}_{s}(R,t)\sum_{i}\biggl(\delta(\textbf{r}-\textbf{r}_{i})\delta(\textbf{p}-\widehat{\textbf{p}}_{i})$$
\begin{equation}\label{SUSDKin def distribution function el subspecies} +\delta(\textbf{p}-\widehat{\textbf{p}}_{i})\delta(\textbf{r}-\textbf{r}_{i})\biggr)\Psi_{s}(R,t)+h.c.\Biggr)dR,\end{equation}
for each subspecies of electrons.

In formula (\ref{SUSDKin def distribution function el subspecies}) we have applied functions $\Psi_{S}(R,t)$, which are the upper $\Psi_{\uparrow}(R,t)$, or lower $\Psi_{\downarrow}(R,t)$ line in the N-rank spinor $\Psi_{S}(R,t)=\left(
                                                                                                     \begin{array}{c}
                                                                                                       \Psi_{\uparrow}(R,t) \\
                                                                                                       \Psi_{\downarrow}(R,t) \\
                                                                                                     \end{array}
                                                                                                   \right).
$

\section{Set of kinetic equations}

We consider quantum plasmas as the set of three species of particles: spin-up electrons, spin-down electrons and ions. Hence we have three kinetic equations presented below.

Kinetic equation for spin-up electrons is
$$\partial_{t}f_{e\uparrow}+\textbf{v}\nabla_{\textbf{r}}f_{e\uparrow} +q_{e}\biggl(\textbf{E}+\frac{1}{c}[\textbf{v},\textbf{B}]\biggr)\nabla_{\textbf{p}}f_{e\uparrow}$$
$$+\gamma_{e}\nabla B^{z} \cdot\nabla_{\textbf{p}} f_{e\uparrow} +\frac{\gamma_{e}}{2}\biggl(\nabla B^{x} \cdot\nabla_{\textbf{p}} S_{e,x}$$
\begin{equation}\label{SUSDKin kin eq f e up}  +\nabla B^{y} \cdot\nabla_{\textbf{p}} S_{e,y}\biggr)=\frac{\gamma_{a}}{\hbar}[S_{e,x} B_{y}-S_{e,y} B_{x}] .\end{equation}

Kinetic equation for spin-down electrons has same structure  as equation for spin-up electrons, but it has some different coefficients
$$\partial_{t}f_{e\downarrow} +\textbf{v}\nabla_{\textbf{r}}f_{e\downarrow}+q_{e}\biggl(\textbf{E}+\frac{1}{c}[\textbf{v},\textbf{B}]\biggr)\nabla_{\textbf{p}}f_{e\downarrow}$$
$$-\gamma_{e}\nabla B^{z}\cdot\nabla_{\textbf{p}} f_{e\downarrow}      +\frac{\gamma_{e}}{2}\biggl(\nabla B^{x} \cdot\nabla_{\textbf{p}} S_{e,x}                  $$
\begin{equation}\label{SUSDKin kin eq f e down} +\nabla B^{y} \cdot\nabla_{\textbf{p}} S_{e,y}\biggr)=-\frac{\gamma_{a}}{\hbar}[S_{e,x} B_{y}-S_{e,y} B_{x}]. \end{equation}

Kinetic equation for ions is
\begin{equation}\label{SUSDKin kin eq f i} \partial_{t}f_{i}+\textbf{v}\nabla_{\textbf{r}}f_{i} +q_{i}\biggl(\textbf{E}+\frac{1}{c}[\textbf{v},\textbf{B}]\biggr)\nabla_{\textbf{p}}f_{i}=0, \end{equation}
where we consider the charge-charge interaction only.

Considering the charge-charge and the spin-spin interactions we apply the self-consistent field approximation \cite{Andreev 1407 Review}, \cite{Andreev IJMP 12}, \cite{Klimontovich book}. The MPQHD equations beyond the self-consistent field approximation can be found in Refs. \cite{MaksimovTMP 2001 b}, \cite{Andreev AtPhys 08}, \cite{Andreev 1403 exchange}.

Kinetic equations for electrons contain spin-distribution functions $S_{e,x}(\textbf{r}, \textbf{p},t)$ and $S_{e,y}(\textbf{r}, \textbf{p},t)$.

The spin distribution functions for each species appears as the quantum mechanical average of the corresponding operator
\begin{equation}\label{SUSDKin operator S} \hat{S}^{\alpha}=\sum_{i}\delta(\textbf{r}-\widehat{\textbf{r}}_{i})\delta(\textbf{p}-\widehat{\textbf{p}}_{i})\widehat{\sigma}^{\alpha}_{i}.\end{equation}
Hence explicit form of the spin distribution function is
$$S_{a}^{\alpha}(\textbf{r}, \textbf{p},t)=\frac{1}{4}\int \Biggl(\Psi_{S}^{+}(R,t)\sum_{i}\biggl(\delta(\textbf{r}-\textbf{r}_{i})\delta(\textbf{p}-\widehat{\textbf{p}}_{i})$$
\begin{equation}\label{SUSDKin_QP def spin distribution function}+\delta(\textbf{p}-\widehat{\textbf{p}}_{i})\delta(\textbf{r}-\textbf{r}_{i})\biggr)\widehat{\sigma}^{\alpha}_{i}\Psi_{S}(R,t)+h.c.\Biggr)dR.\end{equation}
The spin distribution functions $S_{e,x}(\textbf{r}, \textbf{p},t)$ and $S_{e,y}(\textbf{r}, \textbf{p},t)$ appear for all electrons inspite the separation of spin-up and spin-down electrons in the distribution functions.

Differentiating explicit forms of $S_{x}$ and $S_{y}$ and applying the Pauli equation (\ref{SUSDKin Pauli Expl +}) and (\ref{SUSDKin Pauli Expl -}) for the time derivatives of the N-rank spinor wave function $\Psi_{S}$ we obtain the following equations for spin-distribution functions of electrons
$$\partial_{t}S_{e,x}+\textbf{v}\nabla_{\textbf{r}}S_{e,x}+q_{e}\biggl(\textbf{E}+\frac{1}{c}[\textbf{v},\textbf{B}]\biggr)\nabla_{\textbf{p}}S_{e,x}$$
\begin{equation}\label{SUSDKin kin eq S e x} +\gamma_{e}\nabla B^{x}\nabla_{\textbf{p}} (f_{e\uparrow}+f_{e\downarrow}) -\frac{2\gamma_{e}}{\hbar}\biggl(S_{e,y}B^{z}-(f_{e\uparrow}-f_{e\downarrow})B^{y}\biggr)=0, \end{equation}
and
$$\partial_{t}S_{y}+\textbf{v}\nabla_{\textbf{r}}S_{e,y}+q_{e}\biggl(\textbf{E}+\frac{1}{c}[\textbf{v},\textbf{B}]\biggr)\nabla_{\textbf{p}}S_{e,y}$$
\begin{equation}\label{SUSDKin kin eq S e y} +\gamma_{e}\nabla B^{y}\nabla_{\textbf{p}} (f_{e\uparrow}+f_{e\downarrow}) -\frac{2\gamma_{e}}{\hbar}\biggl((f_{e\uparrow}-f_{e\downarrow})B^{x}-S_{e,x}B^{z}\biggr)=0. \end{equation}

Let us mention that $S_{x}$ and $S_{y}$ do not wear subindexes $\uparrow$ and $\downarrow$. As we can see from definitions of $S_{x}$ and $S_{y}$ they are related to both projections spin-up $\Psi_{\uparrow}$ and spin-down $\Psi_{\downarrow}$. Operators $\widehat{\sigma}^{x}_{i}$ and $\widehat{\sigma}^{y}_{i}$ mixing upper and lower components of N-rank spinor wave function. Whereas $\widehat{\sigma}^{z}_{i}$ do not mix them giving $S_{z}(\textbf{r}, \textbf{p},t)=f_{e\uparrow}(\textbf{r}, \textbf{p},t)-f_{e\downarrow}(\textbf{r}, \textbf{p},t)$. The full distribution of all electrons $f_{e}(\textbf{r}, \textbf{p},t)$ is the sum of distribution functions of spin-up $f_{e\uparrow}(\textbf{r}, \textbf{p},t)$ and spin-down $f_{e\downarrow}(\textbf{r}, \textbf{p},t)$ electrons $f_{e}=f_{e\uparrow}(\textbf{r}, \textbf{p},t)+f_{e\downarrow}(\textbf{r}, \textbf{p},t)$.

Electromagnetic fields in the QHD equations presented above obey the Maxwell equations
\begin{equation}\label{SUSDKin div E} \nabla \textbf{E}=4\pi e\biggl(n_{i}-n_{e\uparrow}-n_{e\downarrow}\biggr),\end{equation}
\begin{equation}\label{SUSDKin div B} \nabla \textbf{B}=0, \end{equation}
\begin{equation}\label{SUSDKin rot E} \nabla\times \textbf{E}=-\frac{1}{c}\partial_{t}\textbf{B},\end{equation}
and
$$\nabla\times \textbf{B}=\frac{1}{c}\partial_{t}\textbf{E}+4\pi\nabla\times \textbf{M}_{e}$$
\begin{equation}\label{SUSDKin rot B}
+\frac{4\pi}{c}(q_{e}\textbf{j}_{e\uparrow}+q_{e}\textbf{j}_{e\downarrow}+q_{i}\textbf{j}_{i}),\end{equation}
where $\textbf{M}_{e}=\{\gamma_{e}\tilde{S}_{ex}, \gamma_{e}\tilde{S}_{ey}, \gamma_{e}(n_{e\uparrow}-n_{e\downarrow})\}$ is the magnetization of electrons in terms of hydrodynamic variables.

Material fields entering the Maxwell equations have the following relations to the distribution functions
\begin{equation}\label{SUSDKin}n_{a}(\textbf{r},t)=\int f_{a}(\textbf{r},\textbf{p},t)d\textbf{p},\end{equation}
\begin{equation}\label{SUSDKin}\textbf{j}_{a}(\textbf{r},t)=\int \frac{\textbf{p}}{m_{a}} f_{a}(\textbf{r},\textbf{p},t)d\textbf{p},\end{equation}
\begin{equation}\label{SUSDKin}\tilde{S}_{ex}(\textbf{r},t)=\int S_{ex}(\textbf{r},\textbf{p},t)d\textbf{p},\end{equation}
and
\begin{equation}\label{SUSDKin}\tilde{S}_{ey}(\textbf{r},t)=\int S_{ey}(\textbf{r},\textbf{p},t)d\textbf{p}.\end{equation}

Let us consider the spin density
\begin{equation}\label{SUSDKin_QP def spin density} \tilde{S}^{\alpha}(\textbf{r},t)=\int dR\sum_{i}\delta(\textbf{r}-\textbf{r}_{i})\psi^{*}(R,t)\widehat{\sigma}^{\alpha}_{i}\psi(R,t),\end{equation}
proportional to the magnetization $\textbf{M}_{a}(\textbf{r},t)$, usually used in the
quantum hydrodynamics \cite{MaksimovTMP 2001}, \cite{Andreev spin current}, and \cite{Andreev IJMP 12}: $\textbf{M}_{a}(\textbf{r},t)=\gamma_{a}\textbf{S}_{a}(\textbf{r},t)$. Next integrating the spin distribution function over the momentum we get the spin density appearing in the quantum hydrodynamic equations \cite{Andreev spin current}, \cite{Andreev IJMP 12}

Here we describe explicit form of hydrodynamic spin density projections $\tilde{S}^{\alpha}(\textbf{r},t)$ for the simple single particle case to give a taste of the spin density structure, which reflects structure of the spin-distribution function. Here we need to represent both components of the spinor wave function as $\psi_{s}=a_{s}e^{\imath\phi_{s}}$. These quantities appear as follows $\tilde{S}_{x}=\psi^{*}\sigma_{x}\psi=\psi_{\downarrow}^{*}\psi_{\uparrow}+\psi_{\uparrow}^{*}\psi_{\downarrow}=2a_{\uparrow}a_{\downarrow}\cos\Delta \phi$, $\tilde{S}_{y}=\psi^{*}\sigma_{y}\psi=\imath(\psi_{\downarrow}^{*}\psi_{\uparrow}-\psi_{\uparrow}^{*}\psi_{\downarrow})=-2a_{\uparrow}a_{\downarrow}\sin\Delta \phi$, $\tilde{S}_{z}=\psi_{\uparrow}^{*}\psi_{\uparrow}-\psi_{\downarrow}^{*}\psi_{\downarrow}=a_{\uparrow}^{2}-a_{\downarrow}^{2}$, where $\Delta \phi=\phi_{\uparrow}-\phi_{\downarrow}$. $\tilde{S}_{x}$, $\tilde{S}_{y}$, and $\tilde{S}_{z}$ appear as mixed combinations of $\psi_{\uparrow}$ and $\psi_{\downarrow}$. These quantities do not related to different species of electrons having different spin direction. $\tilde{S}_{x}$ and $\tilde{S}_{y}$ describe simultaneous evolution of both species.

\section{Linearised set of separated spin-up and spin-down quantum kinetic equations}

Operator $[\textbf{v},\textbf{e}_{z}]\partial_{\textbf{p}}$ can be represented as $\frac{1}{m}\partial_{\varphi}$, where $\varphi$ is the angle in the cylindrical coordinates in momentum space.

In equilibrium the set of kinetic equations (\ref{SUSDKin kin eq f e up}), (\ref{SUSDKin kin eq f e down}), (\ref{SUSDKin kin eq f i}), (\ref{SUSDKin kin eq S e x}), and (\ref{SUSDKin kin eq S e y}) has the following form
\begin{equation}\label{SUSDKin} \begin{array}{ccc}
                                  \partial_{\varphi}f_{0e\uparrow}=0, & \partial_{\varphi}f_{0e\downarrow}=0, & \partial_{\varphi}f_{0i}=0,
                                \end{array}
\end{equation}
\begin{equation}\label{SUSDKin eq spin evol eq x} \partial_{\varphi}S_{0e,x} =S_{0e,y},\end{equation}
and
\begin{equation}\label{SUSDKin eq spin evol eq y} \partial_{\varphi}S_{0e,y} =-S_{0e,x}.\end{equation}
We have included that time and space derivatives of the distribution functions equal to zero. We have also included that electric field in equilibrium equals to zero as well. Equilibrium magnetic field equals to the external field directed parallel to the Z axis $B_{x}=B_{y}=0$.

\subsection{Equilibrium distributions}

We consider degenerate electrons. In absence of the external magnetic field two electrons occupy each quantum state with momentum below the Fermi momentum
\begin{equation}\label{SUSDKin equilib distrib el un pol} f_{0e}=\frac{2}{(2\pi\hbar)^{3}}\Theta(p_{Fe}-p),\end{equation}
where $p_{Fe}=(3\pi^{2}n_{0e})^{\frac{1}{3}}\hbar$.

Distribution (\ref{SUSDKin equilib distrib el un pol}) is a spherically symmetric distribution, which does not contain dependence on angle variables $\theta$, $\varphi$.

If we have degenerate electrons in an external magnetic field when occupation of spin-up and spin-down states are different. Under influence of an external magnetic field part of spin-up electrons change direction and transit to spin-down states. Thus instead of the Fermi step containing fully occupied (two electrons in a state) states, which is the unmodified Fermi step, we have two different Fermi steps for spin-up and spin-down electrons. The Fermi step for spin-up electrons is shorter than the unmodified Fermi step, whereas The Fermi step for spin-down electrons is longer than the unmodified Fermi step.  Equilibrium distribution function for each subspecies of electrons are
\begin{equation}\label{SUSDKin equilib distrib el pol} f_{0s}=\frac{1}{(2\pi\hbar)^{3}}\Theta(p_{Fs}-p),\end{equation}
where $p_{Fs}=(6\pi^{2}n_{0s})^{\frac{1}{3}}\hbar$, and $s=\uparrow$, or $\downarrow$.

Distribution function of all electrons is the sum of $f_{0\uparrow}$ and $f_{0\downarrow}$, hence
\begin{equation}\label{SUSDKin} f_{0e}=\frac{1}{(2\pi\hbar)^{3}}[\theta(p_{F\uparrow}-p)+\theta(p_{F\downarrow}-p)], \end{equation}
which pass into (\ref{SUSDKin equilib distrib el un pol}) at $B_{0}\rightarrow 0$.

We consider two limits for ions: classic low temperature ions and degenerate ions.

For classic ions we consider the Maxwell distribution function for equilibrium distribution
\begin{equation}\label{SUSDKin_QP maxwell distrib}f_{0i}(\textbf{p})=\frac{n_{0i}}{(\sqrt{2\pi m_{i} T_{i}})^{3}}\exp\biggl(-\frac{\textbf{p}^{2}}{2 m_{i} T_{i}}\biggr),\end{equation}
where $T_{i}$ is the temperature of classic ions in units of energy, and
$\textbf{p}=m\textbf{v}$.

For degenerate ions we have
\begin{equation}\label{SUSDKin equilib distrib ion un pol} f_{0i}=\frac{2}{(2\pi\hbar)^{3}}\Theta(p_{Fi}-p).\end{equation}
We neglect change of ion occupation number for magnetised ions.

\begin{figure}
\includegraphics[width=8cm,angle=0]{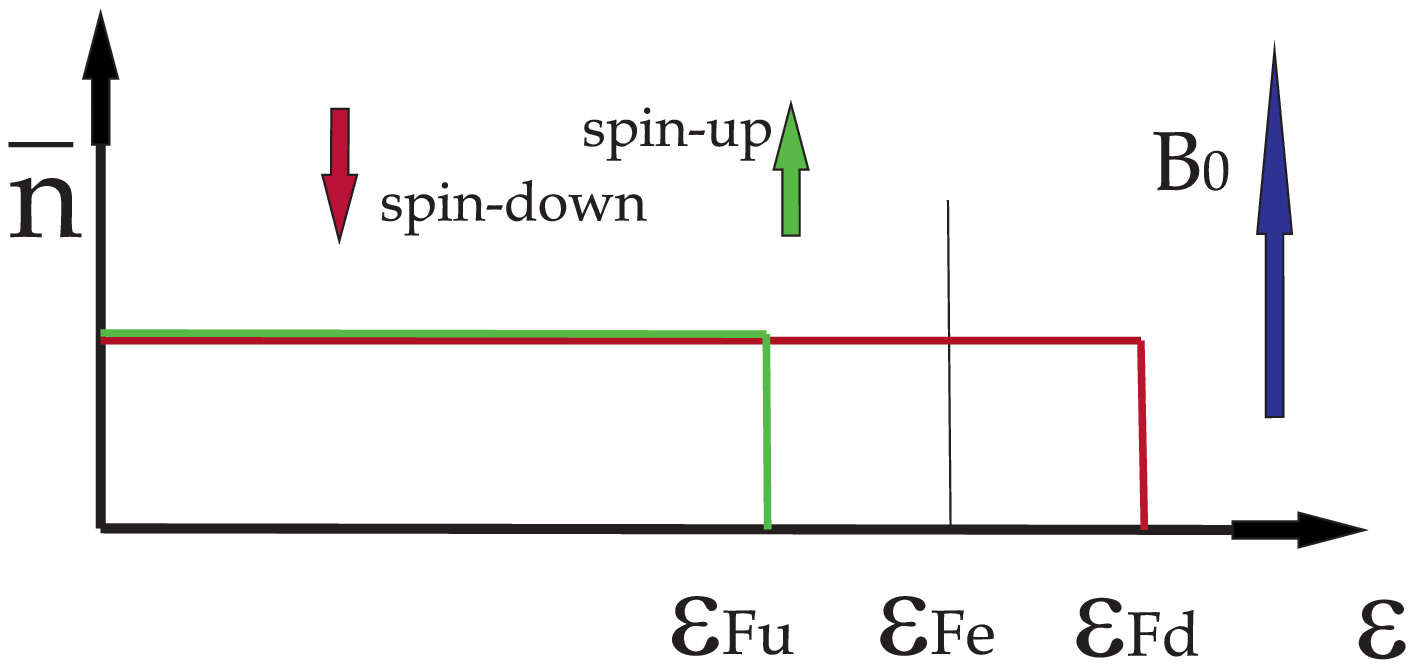}
\includegraphics[width=8cm,angle=0]{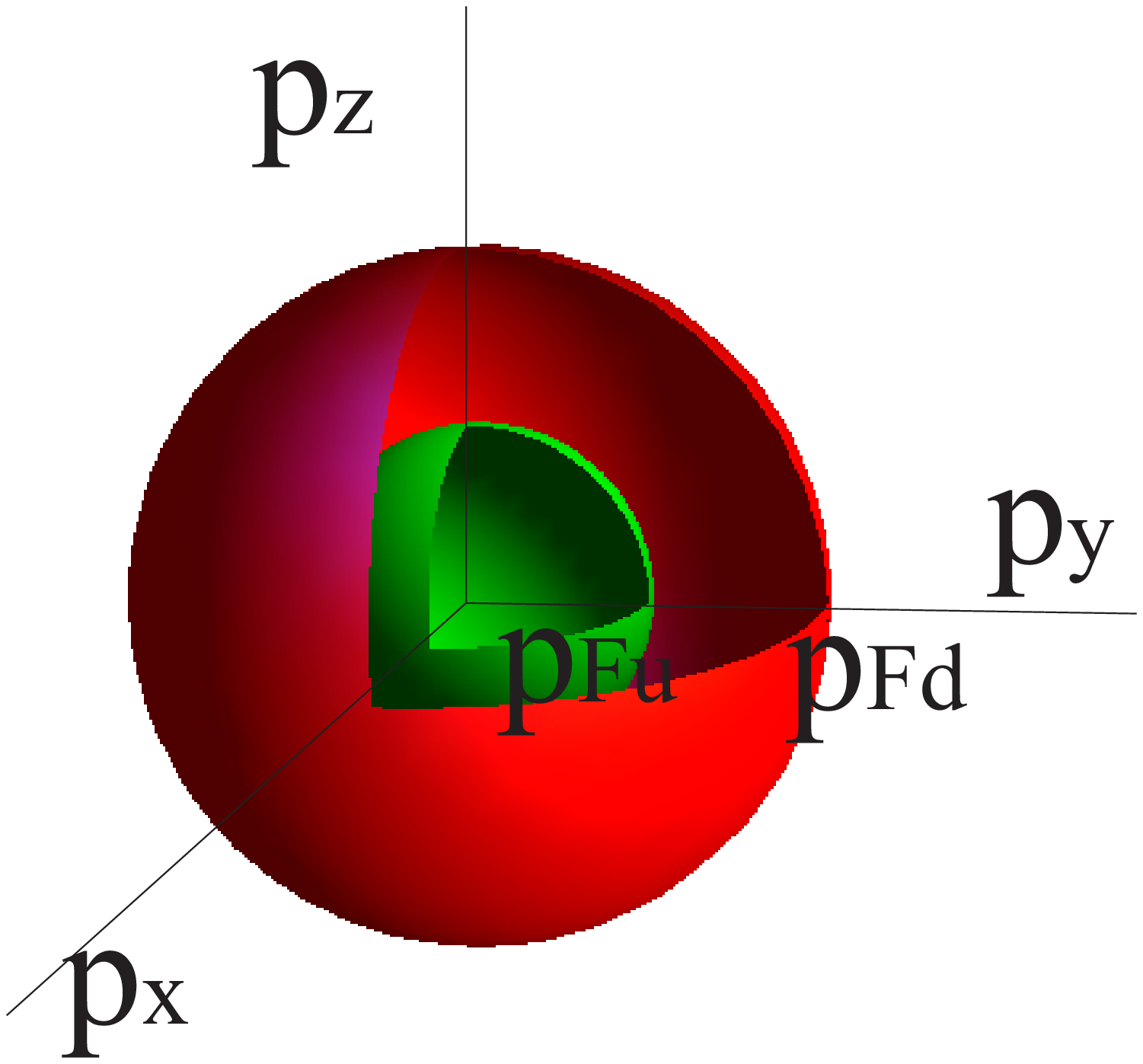}
\caption{\label{SUSDKin F01} (Color online) The figure shows distribution functions $\overline{n}$ of degenerate spin-up and spin-down electrons being in external magnetic field. This distribution function gives average occupation number of quantum states with different energies.}
\end{figure}

From equations (\ref{SUSDKin eq spin evol eq x}) and (\ref{SUSDKin eq spin evol eq y}) we find general form of dependence of equilibrium spin distribution functions on momentum
\begin{equation}\label{SUSDKin equilib distrib spin Gen Str el un pol} \begin{array}{cc}
                                                                                                                       S_{0x}=C(p_{\parallel},p_{\perp})\cos(\varphi+\varphi_{0}), & S_{0y}=C(p_{\parallel},p_{\perp})\sin(\varphi+\varphi_{0}).
                                                                                                                     \end{array}
\end{equation}

Further analysis leads to the following explicit form of equilibrium spin distribution functions
\begin{equation}\label{SUSDKin equilib spin x distrib el un pol} S_{0x}=\frac{1}{(2\pi\hbar)^{3}}\biggl(\Theta(p_{Fu}-p)+\Theta(p_{Fd}-p)\biggr)\cos(\varphi+\varphi_{0}),\end{equation}

\begin{equation}\label{SUSDKin equilib spin ydistrib el un pol} S_{0y}=\frac{1}{(2\pi\hbar)^{3}}\biggl(\Theta(p_{Fu}-p)+\Theta(p_{Fd}-p)\biggr)\sin(\varphi+\varphi_{0}).\end{equation}

Let us mention that corresponding equilibrium hydrodynamic spin densities equal to zero
\begin{equation}\label{SUSDKin} \int S_{0x}(\textbf{p})d\textbf{p}=0,\end{equation}
and
\begin{equation}\label{SUSDKin} \int S_{0y}(\textbf{p})d\textbf{p}=0,\end{equation}
as it should be. These integrals equal to zero due to integration over the angle $\varphi$.

\subsection{Linearised kinetic equations}

Now we are ready to present linearised kinetic equations.

Linearised kinetic equation for spin-up electrons reads as
$$\partial_{t}\delta f_{e\uparrow} +\textbf{v}\partial_{\textbf{r}}\delta f_{e\uparrow}$$ $$+q_{e}\biggl(\delta\textbf{E}+\frac{1}{c}[\textbf{v},\delta\textbf{B}]\biggr)\partial_{\textbf{p}}f_{0e\uparrow} +q_{e}\frac{1}{c}[\textbf{v},\textbf{B}_{0}]\partial_{\textbf{p}}\delta f_{e\uparrow}$$
$$+\gamma_{e}\nabla \delta B^{z} \cdot\nabla_{\textbf{p}} f_{0e\uparrow} +\frac{\gamma_{e}}{2}\biggl(\partial_{\alpha} \delta B^{x} \cdot\partial_{\textbf{p}\alpha} S_{0e,x}$$
\begin{equation}\label{SUSDKin kin eq f e up Lin}   +\nabla \delta B^{y} \cdot\nabla_{\textbf{p}} S_{0e,y}\biggr) =\frac{\gamma_{a}}{\hbar}\biggl(S_{0e,x} \delta B_{y}-S_{0e,y} \delta B_{x}\biggr).\end{equation}

Linearised kinetic equation for spin-down electrons appears as
$$\partial_{t}\delta f_{e\downarrow} +\textbf{v}\nabla_{\textbf{r}}\delta f_{e\downarrow} $$ $$+q_{e}\biggl(\delta\textbf{E}+\frac{1}{c}[\textbf{v},\delta\textbf{B}]\biggr)\nabla_{\textbf{p}}f_{0e\downarrow} +q_{e}\frac{1}{c}[\textbf{v},\textbf{B}_{0}]\nabla_{\textbf{p}}\delta f_{e\downarrow}$$
$$-\gamma_{e}\nabla_{\alpha} \delta B^{z}\cdot\nabla_{\textbf{p}\alpha} f_{0e\downarrow}      +\frac{\gamma_{e}}{2}\biggl(\nabla \delta B^{x} \cdot\nabla_{\textbf{p}} S_{0e,x} $$
\begin{equation}\label{SUSDKin kin eq f e down Lin} +\nabla \delta B^{y} \cdot\nabla_{\textbf{p}} S_{0e,y}\biggr) =-\frac{\gamma_{a}}{\hbar}\biggl(S_{0e,x} \delta B_{y}-S_{0e,y} \delta B_{x}\biggr). \end{equation}

Linearised kinetic equation for ions has the following form
$$\partial_{t}\delta f_{i}+\textbf{v}\nabla_{\textbf{r}}\delta f_{i}$$
\begin{equation}\label{SUSDKin kin eq f i Lin}  +q_{i}\biggl(\delta\textbf{E}+\frac{1}{c}[\textbf{v},\delta\textbf{B}]\biggr)\nabla_{\textbf{p}}f_{0i} +q_{i}\frac{1}{c}[\textbf{v},\textbf{B}_{0}]\nabla_{\textbf{p}}\delta f_{i}=0. \end{equation}

Linearised kinetic equation for x-projection of the spin-distribution function of electrons
$$\partial_{t}\delta S_{e,x}+\textbf{v}\nabla_{\textbf{r}}\delta S_{e,x} $$ $$+q_{e}\biggl(\delta\textbf{E}+\frac{1}{c}[\textbf{v},\delta\textbf{B}]\biggr)\nabla_{\textbf{p}}S_{0e,x} +q_{e}\frac{1}{c}[\textbf{v},\textbf{B}_{0}]\nabla_{\textbf{p}}\delta S_{e,x}$$
$$+\gamma_{e}\nabla \delta B^{x}\cdot\nabla_{\textbf{p}} (f_{0e\uparrow}+f_{0e\downarrow})$$
\begin{equation}\label{SUSDKin kin eq S e x Lin}  =\frac{2\gamma_{e}}{\hbar}\biggl(\delta S_{e,y}B_{0z}+S_{0e,y}\delta B_{z}-(f_{0e\uparrow}-f_{0e\downarrow})\delta B^{y}\biggr), \end{equation}
and linearised kinetic equation for y-projection of the spin-distribution function of electrons
$$\partial_{t}\delta S_{y}+\textbf{v}\nabla_{\textbf{r}}\delta S_{e,y} $$ $$+q_{e}\biggl(\delta\textbf{E}+\frac{1}{c}[\textbf{v},\delta\textbf{B}]\biggr)\nabla_{\textbf{p}}S_{0e,y} +\frac{1}{c}[\textbf{v},\textbf{B}_{0}]\nabla_{\textbf{p}}\delta S_{e,y}$$
$$+\gamma_{e}\nabla \delta B^{y}\cdot\nabla_{\textbf{p}} (f_{0e\uparrow}+f_{0e\downarrow})$$
\begin{equation}\label{SUSDKin kin eq S e y Lin}  =\frac{2\gamma_{e}}{\hbar}\biggl((f_{0e\uparrow}-f_{0e\downarrow})\delta B^{x}-S_{0e,x}\delta B_{z}-\delta S_{e,x} B_{0z}\biggr). \end{equation}

Equations of matter evolution (\ref{SUSDKin kin eq f e up Lin})-(\ref{SUSDKin kin eq S e y Lin}) are coupled with equations of electromagnetic field
\begin{equation}\label{SUSDKin div E Lin} \nabla \delta\textbf{E}=4\pi \sum_{a=u,d,i} q_{a}\int \delta f_{a}(\textbf{r},\textbf{p},t)d\textbf{p},\end{equation}
\begin{equation}\label{SUSDKin div B Lin} \nabla \delta\textbf{B}=0, \end{equation}
\begin{equation}\label{SUSDKin rot E Lin} \nabla\times \delta\textbf{E}=-\frac{1}{c}\partial_{t}\delta\textbf{B},\end{equation}
and
$$\nabla\times \delta\textbf{B}=\frac{1}{c}\partial_{t}\delta\textbf{E}+4\pi\nabla\times \delta\textbf{M}_{e}$$
\begin{equation}\label{SUSDKin rot B Lin}
+\frac{4\pi}{c}\sum_{a=u,d,i} q_{a}\int \frac{\textbf{p}}{m_{a}}\delta f_{a}(\textbf{r},\textbf{p},t)d\textbf{p},\end{equation}
where
\begin{equation}\label{SUSDKin div B Lin}  \delta M_{x}=\gamma_{e}\int \delta S_{e,x}(\textbf{r},\textbf{p},t)d\textbf{p}, \end{equation}
\begin{equation}\label{SUSDKin div B Lin}  \delta M_{y}=\gamma_{e}\int \delta S_{e,y}(\textbf{r},\textbf{p},t)d\textbf{p}, \end{equation}
and
\begin{equation}\label{SUSDKin div B Lin}  \delta M_{z}=\gamma_{e}\int [\delta f_{\uparrow}(\textbf{r},\textbf{p},t)-\delta f_{\downarrow}(\textbf{r},\textbf{p},t)]d\textbf{p}. \end{equation}

\begin{figure}
\includegraphics[width=8cm,angle=0]{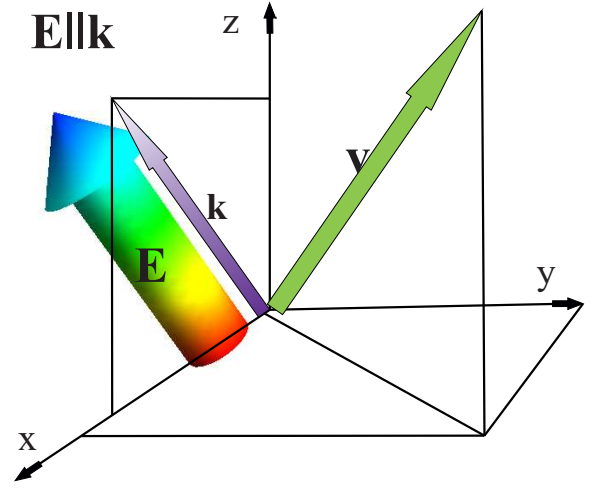}
\caption{\label{SUSDKin F02} (Color online) The figure shows velocity, wave vector, and electric field in oblique propagating longitudinal waves.}
\end{figure}

\subsection{Small amplitude perturbations propagating parallel to the external magnetic field}

Equilibrium condition is described by the non-zero concentrations $n_{0\uparrow}$, $n_{0\downarrow}$, $n_{0}=n_{0\uparrow}+n_{0\downarrow}$, $S_{0x}$, $S_{0y}$, and an external magnetic field $\textbf{B}_{ext}=B_{0}\textbf{e}_{z}$, but $\textbf{E}_{0}=0$.
Assuming that perturbations are monochromatic
\begin{equation}\label{SUSDKin perturbations}
\left(\begin{array}{ccc}
\delta f_{e\uparrow}\\
\delta f_{e\downarrow} \\
\delta f_{i}\\
\delta \textbf{E}\\
\delta \textbf{B}\\
\delta S_{x}\\
\delta S_{y}\\
\end{array}\right)=
\left(\begin{array}{ccc}
F_{A\uparrow} \\
F_{A\downarrow} \\
F_{Ai} \\
\textbf{E}_{A}\\
\textbf{B}_{A}\\
S_{Ax}\\
S_{Ay}\\
\end{array}\right) e^{-\imath\omega t+\imath \textbf{k} \textbf{r}},\end{equation}
we get a set of linear algebraic equations relatively to $F_{A\uparrow}$, $F_{A\downarrow}$, $F_{Ai}$, $V_{A\uparrow}$, $V_{A\downarrow}$, $\textbf{E}_{A}$,
$\textbf{B}_{A}$, $S_{Ax}$, and
$S_{Ay}$. Condition of existence of nonzero solutions for amplitudes of perturbations gives us a dispersion equation.

Difference of spin-up and spin-down concentrations of electrons $\Delta n=n_{0\uparrow}-n_{0\downarrow}$ is caused  by external magnetic field. Since electrons are negative their spins get preferable direction opposite to the external magnetic field $\frac{\Delta n}{n_{0}}=\tanh(\frac{\gamma_{e}B_{0}}{T_{e}})=-\tanh(\frac{\mid\gamma_{e}\mid B_{0}}{T_{e}})$. Here we consider temperature in units of energy, so we do not use the Boltzmann constant.

We consider plasmas in the uniform constant external magnetic field. We see that in linear approach numbers of electrons of each species conserves.

Linearised set of kinetic equations has rather complex form, but we follow results of Refs. \cite{Andreev spin-up and spin-down 1405} and \cite{Andreev spin-up and spin-down 1408 2D}, hence we are focused on properties of the longitudinal waves. This assumption makes analysis more simple. We should mention that waves in magnetised plasmas are not longitudinal in most cases. An exeptional regime supporting propagation of longitudinal waves is limit of wave propagation parallel to external magnetic field. Thus this is the main area of our research. However we obtain an approximate dispersion equation for longitudinal waves in regime of oblique wave propagation.
Longitudinal waves require $\textbf{E}\parallel \textbf{k}$. As a consequence we obtain $\delta \textbf{B}=0$.

\section{Dispersion equation for longitudinal waves: General form}

General form of dispersion equation for oblique propagating longitudinal waves in separated spin evolution model appears as
$$1+\frac{4\pi^{2}e^{2}}{k}\Biggl\{\sum_{s=u,d}\sum_{n=-\infty}^{+\infty}\frac{p_{Fs}^{2}}{(2\pi\hbar)^{3}}\times$$
$$\times\int_{0}^{\pi}\sin\theta d\theta \frac{J_{n}\biggl(\frac{k_{x}v_{Fs}}{\Omega_{s}}\sin\theta\biggr)}{k_{z}v_{Fs}\cos\theta-\omega+n\Omega_{s}}\times$$
$$\times\Biggl[2\cos\alpha\cos\theta J_{n}\biggl(\frac{k_{x}v_{Fs}}{\Omega_{s}}\sin\theta\biggr)$$
$$+\sin\alpha\sin\theta \Biggr(J_{n+1}\biggl(\frac{k_{x}v_{Fs}}{\Omega_{s}}\sin\theta\biggr)+J_{n-1}\biggl(\frac{k_{x}v_{Fs}}{\Omega_{s}}\sin\theta\biggr)\Biggr)\Biggr]$$
$$+\sum_{n=-\infty}^{+\infty}\frac{n_{0i}}{(2\pi m_{i}T_{i})^{\frac{3}{2}}}\frac{1}{2\pi T_{i}}\int d\textbf{p}\frac{J_{n}\biggl(\frac{k_{x}v_{Ti}}{\Omega_{i}}\sin\theta\biggr)e^{-\frac{\textbf{p}^{2}}{2m_{i}T_{i}}}}{k_{z}v_{z}-\omega+n\Omega_{i}}\times$$
$$\times\Biggl(2J_{n}\biggl(\frac{k_{x}v_{Ti}}{\Omega_{i}}\sin\theta\biggr)v_{z}\cos\alpha$$
\begin{equation}\label{SUSDKin}  +\biggl[J_{n+1}\biggl(\frac{k_{x}v_{Ti}}{\Omega_{i}}\sin\theta\biggr)+J_{n-1}\biggl(\frac{k_{x}v_{Ti}}{\Omega_{i}}\sin\theta\biggr)\biggr]\Biggr)\Biggr\}=0, \end{equation}
where $v_{z}=v\cos\theta$, $v_{\perp}=v\sin\theta$, $k_{z}=k\cos\alpha$, $k_{x}=k\sin\alpha$, and $J_{n}(x)$ are the Bessel functions.

Dispersion equation for longitudinal waves propagating parallel to external field in magnetised plasmas rather simpler, in this limit we have
$\textbf{k}\parallel \textbf{B}_{0}$, consequently we obtain $\alpha=0$, $k_{x}=0$, $k_{z}=k$. This assumption leads to more simplification
$J_{n}(0)=0$ if $n\neq0$, and $J_{0}(0)=1$.

After all these simplifications we find
$$1+\frac{8\pi^{2}e^{2}}{k^{2}}\Biggl(\sum_{s=u,d}\frac{m_{e}^{2}v_{Fs}}{(2\pi\hbar)^{3}}\biggl(2+\omega\int_{0}^{\pi}\frac{\sin\theta d\theta}{kv_{Fs}\cos\theta-\omega}\biggr)$$
\begin{equation}\label{SUSDKin DE General for Parallel Pr} +\frac{1}{2\pi T_{i}}\frac{n_{0i}}{(2\pi m_{i}T_{i})^{\frac{1}{2}}}\biggl((2\pi m_{i}T_{i})^{\frac{1}{2}} +\omega\int\frac{e^{-\frac{p_{z}^{2}}{2m_{i}T_{i}}}}{kv_{z}-\omega}dp_{z}\biggr)\Biggr)=0.\end{equation}

For the Maxwell distribution in equilibrium state we meet the following integral in the dispersion equation
$$Z(\alpha)=\frac{1}{\sqrt{\pi}}\int_{-\infty}^{+\infty}\frac{\exp(-\xi^{2})}{\xi-\alpha}d\xi$$
\begin{equation}\label{SUSDKin_QP Z func definition}=\frac{1}{\sqrt{\pi}}\biggl[ P\int_{-\infty}^{+\infty}\frac{\exp(-\xi^{2})}{\xi-\alpha}d\xi\biggr]+\imath\sqrt{\pi}\exp(-\alpha^{2}),\end{equation}
where $\alpha=\frac{\omega}{kv_{T}}$ with $v_{T}\equiv\sqrt{\frac{T}{m}}$, and the symbol $P$ denotes the principle part of the integral.

Let us present assumptions of formula (\ref{SUSDKin_QP Z func definition}).
At $\alpha\gg 1$ we have
\begin{equation}\label{SUSDKin_QP Z big alpha} Z(\alpha)\simeq-\frac{1}{\alpha}\biggl(1+\frac{1}{2\alpha^{2}}+\frac{3}{4\alpha^{4}}+...\biggr)+\imath\sqrt{\pi}\exp(-\alpha^{2}).\end{equation}
This approximate formula will be applied below at description of classic ion contribution in spectrum of plasmas.

\section{Spectrum of longitudinal waves propagating parallel to external field in magnetised separated spin-up and spin-down quantum plasmas}

Taking integrals over angle $\theta$ and $p_{z}$ in equation (\ref{SUSDKin DE General for Parallel Pr}) we get the following explicit form of dispersion equations in regimes of classic and degenerate ions.

\textbf{Classic ions}

Degenerate electrons give, in dispersion equation, well-known logarithmic term. Since we have two species of electrons we obtain two logarithmic terms:
$$1=\frac{3}{2}\frac{\omega_{Lu}^{2}}{v_{Fu}^{2}k^{2}}\biggl(\frac{\omega}{kv_{Fu}}\ln\frac{\omega+kv_{Fu}}{\omega-kv_{Fu}}-2\biggr) $$ $$+\frac{3}{2}\frac{\omega_{Ld}^{2}}{v_{Fd}^{2}k^{2}}\biggl(\frac{\omega}{kv_{Fd}}\ln\frac{\omega+kv_{Fd}}{\omega-kv_{Fd}}-2\biggr) +\frac{\omega_{Li}^{2}}{\omega^{2}}\biggl(1+3\frac{k^{2}v_{Ti}^{2}}{\omega^{2}}\biggr)$$
\begin{equation}\label{SUSDKin DE Longit waves CI}  -\sqrt{\frac{\pi}{2}}\imath\frac{\omega_{Li}^{2}}{k^{2}v_{Ti}^{2}}\frac{\omega}{kv_{Ti}}\exp\biggl(-\frac{\omega^{2}}{2k^{2}v_{Ti}^{2}}\biggr). \end{equation}

\textbf{Quantum degenerate ions}

If ions are degenerate as well as electrons we have three similar logarithmic terms
\begin{equation}\label{SUSDKin  DE Longit waves DI}  1=\sum_{a=u,d,i}\frac{3}{2}\frac{\omega_{La}^{2}}{v_{Fa}^{2}k^{2}}\biggl(\frac{\omega}{kv_{Fa}}\ln\frac{\omega+kv_{Fa}}{\omega-kv_{Fa}}-2\biggr). \end{equation}

Below we present approximate formulas we apply to solve dispersion equations analytically.

At $\omega\gg kv_{a}$ one finds well-known expansion
\begin{equation}\label{SUSDKin} \frac{\omega}{kv_{a}}\ln\frac{\omega+kv_{a}}{\omega-kv_{a}} =2\biggl(1+\frac{1}{3}\frac{k^{2}v_{a}^{2}}{\omega^{2}}+\frac{1}{5}\frac{k^{4}v_{a}^{4}}{\omega^{4}}\biggr). \end{equation}

At $\omega\ll kv_{a}$ we obtain another well-known expansion
\begin{equation}\label{SUSDKin} \frac{\omega}{kv_{a}}\ln\frac{\omega+kv_{a}}{\omega-kv_{a}} =-\pi\imath\frac{\omega}{kv_{a}}+2\frac{\omega^{2}}{k^{2}v_{a}^{2}}\biggl(1+\frac{1}{3}\frac{\omega^{2}}{k^{2}v_{a}^{2}}\biggr). \end{equation}

\subsection{Langmuir wave}

Dispersion equation for high frequency regime at classic ions
$$1= \frac{\omega_{Ld}^{2}}{\omega^{2}}\biggl(1+\frac{3}{5}\frac{k^{2}v_{Fd}^{2}}{\omega^{2}}\biggr) $$
$$+\frac{\omega_{Lu}^{2}}{\omega^{2}}\biggl(1+\frac{3}{5}\frac{k^{2}v_{Fu}^{2}}{\omega^{2}}\biggr) +\frac{\omega_{Li}^{2}}{\omega^{2}}\biggl(1+3\frac{k^{2}v_{Ti}^{2}}{\omega^{2}}\biggr)$$
\begin{equation}\label{SUSDKin LW DE CI}   -\sqrt{\frac{\pi}{2}}\imath\frac{\omega_{Li}^{2}}{k^{2}v_{Ti}^{2}}\frac{\omega}{kv_{Ti}}\exp\biggl(-\frac{\omega^{2}}{2k^{2}v_{Ti}^{2}}\biggr). \end{equation}

Dispersion equation for high frequency regime at classic ions
$$1= \frac{\omega_{Lu}^{2}}{\omega^{2}}\biggl(1+\frac{3}{5}\frac{k^{2}v_{Fu}^{2}}{\omega^{2}}\biggr) $$
\begin{equation}\label{SUSDKin LW DE DI}   +\frac{\omega_{Lu}^{2}}{\omega^{2}}\biggl(1+\frac{3}{5}\frac{k^{2}v_{Fu}^{2}}{\omega^{2}}\biggr) +\frac{\omega_{Li}^{2}}{\omega^{2}}\biggl(1+3\frac{k^{2}v_{Ti}^{2}}{\omega^{2}}\biggr). \end{equation}

In general case the frequency of excitations appears in complex form $\omega=\omega_{R}+\imath\omega_{Im}$.

Spectrums of the Langmuir waves are
$$\omega^{2}_{R}=(\omega^{2}_{Ld}+\omega^{2}_{Lu}+\omega^{2}_{Li})$$
\begin{equation}\label{SUSDKin LW DD DI}  +\frac{3}{5}k^{2}\biggl(\frac{n_{0u}}{n_{0e}}v_{Fu}^{2}+\frac{n_{0d}}{n_{0e}}v_{Fd}^{2}+\frac{m_{e}}{m_{i}}v_{Fi}^{2}\biggr) \end{equation}
for degenerate ions, and
$$\omega^{2}_{R}=(\omega^{2}_{Ld}+\omega^{2}_{Lu}+\omega^{2}_{Li})$$
\begin{equation}\label{SUSDKin LW DD CI}  +\frac{3}{5}k^{2}\biggl(\frac{n_{0u}}{n_{0e}}v_{Fu}^{2}+\frac{n_{0d}}{n_{0e}}v_{Fd}^{2}+\frac{5}{3}\frac{m_{e}}{m_{i}}v_{Ti}^{2}\biggr), \end{equation}
with
\begin{equation}\label{SUSDKin LW LD} \omega_{Im}=\frac{1}{2}\sqrt{\frac{\pi}{2}}\imath\frac{\omega_{Li}^{2}}{k^{2}v_{Ti}^{2}}\frac{\omega^{2}_{Le}}{kv_{Ti}}\exp\biggl(-\frac{\omega^{2}_{Le}}{2k^{2}v_{Ti}^{2}}\biggr)\end{equation}
for classic electrons.

In formulae (\ref{SUSDKin LW DD DI}) and (\ref{SUSDKin LW DD CI}) sum of partial Langmuir frequencies $\omega^{2}_{Ld}+\omega^{2}_{Lu}$ gives full Langmuir frequency of electrons $\omega^{2}_{Le}=\omega^{2}_{Ld}+\omega^{2}_{Lu}$, since $n_{0e}=n_{0u}+n_{0d}$.

The Langmuir wave in degenerate electron gas does not have collisionless damping $\omega_{Im}=0$ if ions are degenerate as well.
In regime of classic (Maxwellian) ions, there is small damping of the Langmuir waves in degenerate electron gas.

In Ref. \cite{Andreev spin-up and spin-down 1405} we have obtained $\omega^{2}=(\omega^{2}_{Ld}+\omega^{2}_{Lu})+\frac{1}{3}k^{2}(\frac{n_{0u}}{n_{0e}}v_{Fu}^{2}+\frac{n_{0d}}{n_{0e}}v_{Fd}^{2})$. We can see that pressure term has different coefficient. This difference appeared due to application of the Fermi pressure (for spin-up and spin-down separately) as an equation of state $P_{Fs}=\frac{1}{5}\frac{(6\pi^{2})^{\frac{2}{3}}n_{s}^{\frac{5}{3}}\hbar^{2}}{m}$. The Fermi pressure gives the equation of state for equilibrium, whereas we considered perturbations of an equilibrium state. To cancel the difference between hydrodynamic and kinetics of small perturbations in three dimensional plasmas we can write down the following modified equation of state $P_{ms}=\frac{1}{5}\frac{mv_{Fs}^{2}}{n_{0s}^{2}}n_{s}^{3}$ (see Ref. \cite{Andreev 1407 Review} formula 99), where $n_{s}$ is the full concentration of particles with $s$ spin projection on z direction, and $n_{0s}$ is the equilibrium concentration of particles with $s$ spin projection. This formula gives the Fermi pressure in equilibrium $n=n_{0}$, and it gives spectrum coinciding with results of kinetic theory of degenerate electron gas.

Let us represent the real part of the Langmuir spectrum in approximate, and more explicit form. We present this spectrum in terms of conventional variables $\omega_{Le}$, $v_{Fe}$, and $\Delta n$, hence we have
\begin{equation}\label{SUSDKin} \omega^{2}=\omega^{2}_{Le}+\frac{3}{10}k^{2}v_{Fe}^{2}\biggl[\biggl(1-\frac{\Delta n}{n_{0e}}\biggr)^{\frac{5}{3}} +\biggl(1+\frac{\Delta n}{n_{0e}}\biggr)^{\frac{5}{3}}\biggr]. \end{equation}
This dependence on $\Delta n$ corresponds to results obtained in Refs. \cite{MaksimovTMP 2001 b}, \cite{Andreev 1403 exchange}.

\subsection{Spin-electron acoustic waves}

In this subsection we present one of main results of this paper. We present the kinetic analysis of the spin-electron acoustic waves. At hydrodynamic description we were able to get spectrum of the SEAWs at all wave vectors \cite{Andreev spin-up and spin-down 1405}, \cite{Andreev spin-up and spin-down 1406 Oblique}. Here we can get an analytical solution for intermediate frequencies described below (see conditions (\ref{SUSDKin cond SEAW CI}) and (\ref{SUSDKin SEAW DI Cond})). Nevertheless, the quantum kinetics allows us to study the Landau damping of the SEAWs.

Regime of high spin polarisation allows to perform analytic consideration of the spin-electron acoustic wave spectrum.

\subsubsection{SEAW: Classic ions}

Part of spectrum of spin-electron acoustic wave can be derived at the following conditions
\begin{equation}\label{SUSDKin cond SEAW CI} kv_{Ti}, kv_{Fu}\ll\omega\ll kv_{Fd}. \end{equation}

Dispersion equation (\ref{SUSDKin DE Longit waves CI}) simplifies at conditions (\ref{SUSDKin cond SEAW CI}). Its simple form appears as
$$1+3 \frac{\omega_{Ld}^{2}}{k^{2}v_{Fd}^{2}}\biggl(1+\frac{\pi}{2}\imath\frac{\omega}{kv_{Fd}}-\frac{\omega^{2}}{k^{2}v_{Fd}^{2}}\biggr)$$

$$ = \frac{\omega_{Lu}^{2}}{\omega^{2}}\biggl(1+\frac{3}{5}\frac{k^{2}v_{Fu}^{2}}{\omega^{2}}\biggr) +\frac{\omega_{Li}^{2}}{\omega^{2}}\biggl(1+3\frac{k^{2}v_{Ti}^{2}}{\omega^{2}}\biggr)$$
\begin{equation}\label{SUSDKin DE SEAW CI} -\sqrt{\frac{\pi}{2}}\imath\frac{\omega_{Li}^{2}}{k^{2}v_{Ti}^{2}}\frac{\omega}{kv_{Ti}}\exp\biggl(-\frac{\omega^{2}}{2k^{2}v_{Ti}^{2}}\biggr). \end{equation}

In major order equation (\ref{SUSDKin DE SEAW CI}) gives the following spectrum
\begin{equation}\label{SUSDKin SEAW DD I order} \omega_{R0}^{2}=\frac{\biggl(\omega^{2}_{Lu}+\omega^{2}_{Li}\biggr)}{1+3\frac{\omega_{Ld}^{2}}{k^{2}v_{Fd}^{2}}}. \end{equation}

Including terms of second order we obtain more general dispersion dependence
\begin{equation}\label{SUSDKin SEAW DD II order} \omega_{R}^{2} =\frac{\omega^{2}_{Lu}\biggl(1+\frac{3}{5}\frac{k^{2}v_{Fu}^{2}}{\omega_{R0}^{2}}\biggr) +\omega^{2}_{Li}\biggl(1+3\frac{k^{2}v_{Ti}^{2}}{\omega_{R0}^{2}}\biggr)}{1+3\frac{\omega_{Ld}^{2}}{k^{2}v_{Fd}^{2}}-3\frac{\omega_{Ld}^{2}\omega_{R0}^{2}}{k^{4}v_{Fd}^{4}}}. \end{equation}

We also obtain imaginary part of the frequency giving Landau damping of the SEAW
\begin{equation}\label{SUSDKin SEAW LD} \omega_{Im} =\frac{1}{2}\omega_{R}\frac{\frac{3\pi}{2}\frac{\omega_{Ld}^{2}}{k^{2}v_{Fd}^{2}}\frac{\omega_{R0}}{kv_{Fd}} +\sqrt{\frac{\pi}{2}}\frac{\omega_{Li}^{2}}{k^{2}v_{Ti}^{2}}\frac{\omega_{R0}}{kv_{Ti}}\exp\biggl(-\frac{\omega^{2}_{R0}}{2k^{2}v_{Ti}^{2}}\biggr)}{1+3\frac{\omega_{Ld}^{2}}{k^{2}v_{Fd}^{2}}-3\frac{\omega_{Ld}^{2}\omega_{R0}^{2}}{k^{4}v_{Fd}^{4}}}.\end{equation}

In long-wavelength regime $\omega_{Ld}\gg kv_{Fd}$ formula (\ref{SUSDKin SEAW DD I order}) simplifies to
\begin{equation}\label{SUSDKin SEAW DD LWLR} \omega_{R0}^{2}=\frac{1}{3}k^{2}v_{Fd}^{2}\frac{(\omega^{2}_{Lu}+\omega^{2}_{Li})}{\omega^{2}_{Ld}}. \end{equation}

At intermediate spin polarisation $\omega^{2}_{Lu}\gg\omega^{2}_{Li}$ we can neglect ion contribution in formula (\ref{SUSDKin SEAW DD LWLR}) and find
\begin{equation}\label{SUSDKin} \omega_{R0}^{2}=\frac{1}{3}\frac{n_{0u}}{n_{0d}}k^{2}v_{Fd}^{2} .\end{equation}

\subsubsection{SEAW: Degenerate ions}

Dispersion equation for the SEAW has form of
$$1+3 \frac{\omega_{Ld}^{2}}{k^{2}v_{Fd}^{2}}\biggl(1+\frac{\pi}{2}\imath\frac{\omega}{kv_{Fd}}-\frac{\omega^{2}}{k^{2}v_{Fd}^{2}}\biggr)$$
\begin{equation}\label{SUSDKin SEAW DI DE} = \frac{\omega_{Lu}^{2}}{\omega^{2}}\biggl(1+\frac{3}{5}\frac{k^{2}v_{Fu}^{2}}{\omega^{2}}\biggr) +\frac{\omega_{Li}^{2}}{\omega^{2}}\biggl(1+\frac{3}{5}\frac{k^{2}v_{Fi}^{2}}{\omega^{2}}\biggr). \end{equation}

Equation (\ref{SUSDKin SEAW DI DE}) arises at
conditions
\begin{equation}\label{SUSDKin SEAW DI Cond} kv_{Fi}, kv_{Fu}\ll\omega\ll kv_{Fd}. \end{equation}

Equation (\ref{SUSDKin SEAW DI DE}) gives spectrum of the SEAWs
\begin{equation}\label{SUSDKin SEAW DI Spectr I order} \omega_{R0}^{2}=\frac{\biggl(\omega^{2}_{Lu}+\omega^{2}_{Li}\biggr)}{1+3\frac{\omega_{Ld}^{2}}{k^{2}v_{Fd}^{2}}}. \end{equation}

Landau damping of the SEAW is found to be
\begin{equation}\label{SUSDKin SEAW DI LD} \omega_{Im}=\frac{1}{2}\omega_{R} \frac{\frac{3\pi}{2}\frac{\omega_{Ld}^{2}}{k^{2}v_{Fd}^{2}}\frac{\omega_{R0}}{kv_{Fd}}}{1+3\frac{\omega_{Ld}^{2}}{k^{2}v_{Fd}^{2}}-3\frac{\omega_{Ld}^{2}\omega_{R0}^{2}}{k^{4}v_{Fd}^{4}}}. \end{equation}

At $\omega_{Ld}^{2}\gg k^{2}v_{Fd}^{2}$ we find simplification of formula (\ref{SUSDKin SEAW DI LD}) $\omega_{Im}=\frac{\pi}{4}\frac{\omega_{R0}}{kv_{Fd}}\omega_{R0}\ll\omega_{R0}$.

In opposite limit $\omega_{Ld}^{2}\ll k^{2}v_{Fd}^{2}$ the denominator in formula (\ref{SUSDKin SEAW DI LD}) equals to 1 and we have $\omega_{Im}=\frac{3\pi}{4}\frac{\omega_{Ld}^{2}}{k^{2}v_{Fd}^{2}}\frac{\omega_{R0}}{kv_{Fd}}\omega_{R0}\ll\omega_{R0}$.

So, the Landau damping of the SEAWs always smaller than frequency of the wave. Thus we have found that the SEAW is a weakly damped wave.

Including smaller corrections in equation (\ref{SUSDKin SEAW DI DE}) we can find generalisation of formula (\ref{SUSDKin SEAW DI Spectr I order})
\begin{equation}\label{SUSDKin SEAW DI Spectr II order} \omega_{R}^{2}=\frac{\omega^{2}_{Lu}\biggl(1+\frac{3}{5}\frac{k^{2}v_{Fu}^{2}}{\omega_{R0}^{2}}\biggr) +\omega^{2}_{Li}\biggl(1+\frac{3}{5}\frac{k^{2}v_{Fi}^{2}}{\omega_{R0}^{2}}\biggr)}{1+3\frac{\omega_{Ld}^{2}}{k^{2}v_{Fd}^{2}}-3\frac{\omega_{Ld}^{2}\omega_{R0}^{2}}{k^{4}v_{Fd}^{4}}}. \end{equation}

Formula (\ref{SUSDKin SEAW DD I order}) coincides with the result for Maxwellian ions (\ref{SUSDKin SEAW DD I order}).
Hence its long-wavelength limit ($\omega_{Ld}\gg kv_{Fd}$) coincides with formula (\ref{SUSDKin SEAW DD LWLR}).

\subsubsection{SEAW: Discussion}

Formulae (\ref{SUSDKin SEAW DD I order}) and (\ref{SUSDKin SEAW DI Spectr I order}) can be rewritten in terms of $\omega_{Le}$ and $v_{Fe}$. This representation explicitly shows contribution of mismatch of the Fermi surfaces of spin-up and spin-down electrons.
\begin{equation}\label{SUSDKin SEAW DD CI CON} \omega_{R0}^{2}=\frac{\frac{1}{2}(1-\frac{\Delta n}{n_{0e}})\omega^{2}_{Le}+\omega^{2}_{Li}}{1+\frac{3}{2}\frac{\omega_{Le}^{2}}{k^{2}v_{Fe}^{2}}(1+\frac{\Delta n}{n_{0e}})^{\frac{1}{3}}}. \end{equation}

At $\omega_{Ld}\gg kv_{Fd}$
and $\omega^{2}_{Lu}\gg\omega^{2}_{Li}$ formula (\ref{SUSDKin SEAW DD CI CON}) simplifies and we find
\begin{equation}\label{SUSDKin SEAW DD CI CON limit case} \omega_{R0}^{2}=\frac{1}{3}\frac{(1-\frac{\Delta n}{n_{0e}})}{(1+\frac{\Delta n}{n_{0e}})^{\frac{1}{3}}}\cdot k^{2}v_{Fe}^{2} .\end{equation}

In this subsection we work under condition $v_{Fd}\gg v_{Fu}$ $\Longrightarrow$ $2>1+\frac{\Delta n}{n_{0e}}\gg 1+\frac{\Delta n}{n_{0e}}$ $\Longrightarrow$ $\Delta n$ comparable with $n_{0}$. Thus we conclude that phase velocity of the SEAW given by formula (\ref{SUSDKin SEAW DD CI CON limit case}) considerably less than the electron Fermi velocity $\omega_{R0}\approx\frac{1}{\sqrt{6}}\sqrt{1-\frac{\Delta n}{n_{0e}}}kv_{Fe}\ll kv_{Fe}$.

\subsection{SEAW: Regime of phase velocity near the Fermi velocity of spin-up electrons $v_{Fu}$}

Let us consider the limit $\omega\rightarrow kv_{Fu}$, which is the low frequency analog of the zeroth sound. In this case we can present frequency of oscillations as $\omega=kv_{Fu}+\delta\omega$. Dispersion equation arises as
$$1=\frac{3}{2}\frac{\omega_{Lu}^{2}}{k^{2}v_{Fu}^{2}}\biggl[\frac{\omega\mid_{\omega\approx kv_{Fu}}}{kv_{Fu}} \ln\biggl(\frac{2kv_{Fu}}{\delta\omega}\biggr)-2\biggr]$$
\begin{equation}\label{SUSDKin LoW Fr ZeroTh sound analog of SEAW} -\frac{3}{2}\frac{\omega_{Ld}^{2}}{k^{2}v_{Fd}^{2}}\biggl(2 +\frac{n_{0u}^{\frac{1}{3}}}{n_{0d}^{\frac{1}{3}}}\ln\frac{v_{Fd}-v_{Fu}}{v_{Fd}+v_{Fu}} +\pi\imath\biggl(\frac{n_{0u}}{n_{0d}}\biggr)^{\frac{1}{3}}\biggr). \end{equation}

Corresponding dispersion dependence arises as
$$\omega=kv_{Fu}\biggl\{1-2\frac{v_{Fd}+v_{Fu}}{v_{Fd}-v_{Fu}}\times$$
\begin{equation}\label{SUSDKin}  \times\exp\biggl[-2 -\frac{2}{3}\frac{k^{2}v_{Fu}^{2}}{\omega_{Lu}^{2}} -2\biggl(\frac{n_{0d}}{n_{0u}}\biggr)^{\frac{1}{3}} \biggr]\biggr\} .\end{equation}

\subsection{Ion-acoustic wave}

\subsubsection{Classic ions}

Ion-acoustic waves exist at the following conditions
\begin{equation}\label{SUSDKin} kv_{Ti}\ll\omega\ll kv_{Fu}, kv_{Fd}. \end{equation}

Dispersion equation for ion acoustic waves with classic ions has the following form
$$1+3\biggl(\frac{\omega_{Ld}^{2}}{k^{2}v_{Fd}^{2}} +\frac{\omega_{Lu}^{2}}{k^{2}v_{Fu}^{2}}\biggr)$$

$$+\frac{3}{2}\pi\imath\omega\biggl(\frac{\omega_{Ld}^{2}}{k^{3}v_{Fd}^{3}} +\frac{\omega_{Lu}^{2}}{k^{3}v_{Fu}^{3}}\biggr)$$
\begin{equation}\label{SUSDKin DE IAW CI} =\frac{\omega_{Li}^{2}}{\omega^{2}} -\sqrt{\frac{\pi}{2}}\imath\frac{\omega_{Li}^{2}}{k^{2}v_{Ti}^{2}}\frac{\omega}{kv_{Ti}}\exp\biggl(-\frac{\omega^{2}}{2k^{2}v_{Ti}^{2}}\biggr). \end{equation}

Equation (\ref{SUSDKin DE IAW CI}) gives dispersion dependence of ion-acoustic waves
\begin{equation}\label{SUSDKin DD IAW CI} \omega_{R}^{2} =\frac{\omega_{Li}^{2}}{1+3(\frac{\omega_{Ld}^{2}}{k^{2}v_{Fd}^{2}} +\frac{\omega_{Lu}^{2}}{k^{2}v_{Fu}^{2}})}. \end{equation}

If $\omega_{Ls}\gg kv_{Fs}$ we find long-wavelength limit of dispersion dependence (\ref{SUSDKin DD IAW CI})
\begin{equation}\label{SUSDKin} \omega_{R}^{2} =\frac{1}{3}k^{2} \frac{v_{Fd}^{2}v_{Fu}^{2}\omega_{Li}^{2}}{v_{Fd}^{2}\omega_{Lu}^{2}+v_{Fu}^{2}\omega_{Ld}^{2}}. \end{equation}

Imaginary part of frequency of ion-acoustic waves at Maxwellian ion appears as
$$\omega_{Im}=-\frac{3\pi}{4}\frac{\omega_{R}^{4}}{\omega_{Li}^{2}} \biggl(\frac{\omega_{Ld}^{2}}{k^{3}v_{Fd}^{3}} +\frac{\omega_{Lu}^{2}}{k^{3}v_{Fu}^{3}}\biggr)$$
\begin{equation}\label{SUSDKin} -\frac{1}{2}\sqrt{\frac{\pi}{2}}\biggl(\frac{\omega_{R}}{kv_{Ti}}\biggr)^{3}\omega_{R}\exp\biggl(-\frac{\omega_{R}^{2}}{2k^{2}v_{Ti}^{2}}\biggr).  \end{equation}

\subsubsection{Degenerate ions}

Conditions of existence of the ion-acoustic waves in plasmas of degenerate electrons and ions are
\begin{equation}\label{SUSDKin} kv_{Fi}\ll\omega\ll kv_{Fu}, kv_{Fd} .\end{equation}

In this regime we can obtain the dispersion equation
$$1+3\biggl(\frac{\omega_{Ld}^{2}}{k^{2}v_{Fd}^{2}} +\frac{\omega_{Lu}^{2}}{k^{2}v_{Fu}^{2}}\biggr)$$

\begin{equation}\label{SUSDKin DE IAw DI} +\frac{3}{2}\pi\imath\omega\biggl(\frac{\omega_{Ld}^{2}}{k^{3}v_{Fd}^{3}} +\frac{\omega_{Lu}^{2}}{k^{3}v_{Fu}^{3}}\biggr)=\frac{\omega_{Li}^{2}}{\omega^{2}}. \end{equation}

Equation (\ref{SUSDKin DE IAw DI}) gives the following solution in leading order on small parameters
\begin{equation}\label{SUSDKin DD IAw DI 0th order} \omega_{R}^{2} =\frac{\omega_{Li}^{2}}{1+3(\frac{\omega_{Ld}^{2}}{k^{2}v_{Fd}^{2}} +\frac{\omega_{Lu}^{2}}{k^{2}v_{Fu}^{2}})}. \end{equation}

At $\omega_{Ls}\gg kv_{Fs}$ (long-wavelength regime) we find simplification of solution (\ref{SUSDKin DD IAw DI 0th order}) as follows
\begin{equation}\label{SUSDKin DD IAw DI oth Lwl limit} \omega_{R}^{2} =\frac{1}{3}k^{2} \frac{v_{Fd}^{2}v_{Fu}^{2}\omega_{Li}^{2}}{v_{Fd}^{2}\omega_{Lu}^{2}+v_{Fu}^{2}\omega_{Ld}^{2}}. \end{equation}

Including smaller corrections to the spectrum (\ref{SUSDKin DD IAw DI 0th order}) find decrement of the Landau damping for ion-acoustic waves
\begin{equation}\label{SUSDKin DD IAw DI LD} \omega_{Im}=-\frac{3\pi}{4}\frac{\omega_{R}^{4}}{\omega_{Li}^{2}} \biggl(\frac{\omega_{Ld}^{2}}{k^{3}v_{Fd}^{3}} +\frac{\omega_{Lu}^{2}}{k^{3}v_{Fu}^{3}}\biggr). \end{equation}

\subsubsection{Ion acoustic waves: Discussion}

At $kv_{Fs}\gg\omega_{Ls}$ formula (\ref{SUSDKin DD IAw DI 0th order})  gives well-known limit $\omega_{R}^{2}=\omega_{Li}^{2}$.

Next let us consider formula (\ref{SUSDKin DD IAw DI oth Lwl limit}), which has been obtained from formula (\ref{SUSDKin DD IAw DI 0th order}) in regime of long-wavelength $\omega_{Ls}\gg kv_{Fs}$, in more explicit form
\begin{equation}\label{SUSDKin} \omega_{R}^{2}=\frac{2}{3}\frac{m_{e}}{m_{i}}k^{2}v_{Fe}^{2}\frac{1}{(1-\frac{\Delta n}{n_{0e}})^{\frac{1}{3}}+(1+\frac{\Delta n}{n_{0e}})^{\frac{1}{3}}}. \end{equation}
If magnetic field is small than $\Delta n/n_{0e}\ll1$. In this regime we obtain
\begin{equation}\label{SUSDKin} \omega_{R}^{2}=\frac{1}{3}\frac{m_{e}}{m_{i}}k^{2}v_{Fe}^{2} \biggl(1+\frac{1}{9}\frac{\Delta n^{2}}{n_{0e}^{2}}\biggr). \end{equation}

In the long-wavelength limit $\omega_{Ls}\gg kv_{Fs}$ the decrement of Landau damping for ion acoustic wave (\ref{SUSDKin DD IAw DI LD}) can be rewritten as
\begin{equation}\label{SUSDKin DD IAw DI LD Lwl limit} \omega_{Im}=-\frac{\pi}{12}\frac{m_{e}}{m_{i}}kv_{Fe}\frac{1}{[(1-\frac{\Delta n}{n_{0e}})^{\frac{1}{3}}+(1+\frac{\Delta n}{n_{0e}})^{\frac{1}{3}}]^{2}}. \end{equation}

We can find simplification of formula (\ref{SUSDKin DD IAw DI LD Lwl limit}) for regime of small magnetic field
\begin{equation}\label{SUSDKin DD IAw DI LD Lwl limit   small B} \omega_{Im}=-\frac{\pi}{48}\frac{m_{e}}{m_{i}}kv_{Fe}\biggl(1+\frac{2}{9}\frac{\Delta n^{2}}{n_{0e}^{2}}\biggr). \end{equation}

\subsection{Zeroth sound}

The zeroth sound is a well-known high frequency solution of the dispersion equation for degenerate ions. Usually one obtains it for an equal occupation of spin-up and spin-down states. Now we consider the zeroth sound in regime of high difference in occupation of spin-up and spin-down states by degenerate electrons. In this case  $\omega\sim kv_{Fd}\gg kv_{Fu}\gg kv_{Fi}$.

The zero-sound (Ref on Silin) appears at $\omega\rightarrow kv_{Fd}$
\begin{equation}\label{SUSDKin} \omega=kv_{Fd}+\delta\omega, \end{equation}
where $\delta\omega\ll kv_{Fd}$

In this regime dispersion equation takes the following form
\begin{equation}\label{SUSDKin DE ZeroTh sound} 1=\frac{3}{2}\frac{\omega_{Ld}^{2}}{k^{2}v_{Fd}^{2}}\biggl[\frac{\omega\mid_{\omega\approx kv_{Fd}}}{kv_{Fd}} \ln\biggl(\frac{2kv_{Fd}}{\delta\omega}\biggr)-2\biggr]. \end{equation}

Equation (\ref{SUSDKin DE ZeroTh sound}) gives the following solution
\begin{equation}\label{SUSDKin DD Zero s} \omega_{0}=kv_{Fd}+\delta\omega=kv_{Fd}\biggl[1+2\exp\biggl(-2-\frac{2}{3}\frac{k^{2}v_{Fd}^{2}}{\omega_{Ld}^{2}}\biggr)\biggr]. \end{equation}

Let us present here the Fermi velocity of spin-down electrons via the conventional Fermi velocity
\begin{equation}\label{SUSDKin VFd via VFe} v_{Fd}=v_{Fe}\biggl(1+\frac{\Delta n}{n_{0e}}\biggr)^{\frac{1}{3}}, \end{equation}
with $v_{fe}=(3\pi^{2}n_{0e})^{\frac{1}{3}}\hbar/m$.

More explicit form of the zeroth sound spectrum (\ref{SUSDKin DD Zero s}) appears at substitution (\ref{SUSDKin VFd via VFe}) in formula (\ref{SUSDKin DD Zero s}). Hence we have
$$\omega_{0}=kv_{Fe}\biggl(1+\frac{\Delta n}{n_{0e}}\biggr)^{\frac{1}{3}}\times$$
\begin{equation}\label{SUSDKin} \times\biggl[1+2\exp\biggl(-2-\frac{4k^{2}v_{Fe}^{2}}{3\omega_{Ld}^{2}}\biggl(1+\frac{\Delta n}{n_{0e}}\biggr)^{-\frac{1}{3}}\biggr)\biggr]. \end{equation}

We can also consider regime of small difference in occupation numbers as well. In this case $kv_{Fd}\sim kv_{Fu}$, and $\omega\sim kv_{Fd}> kv_{Fu}\gg kv_{Fi}$.

\begin{equation}\label{SUSDKin} \omega=kv_{Fd}+\delta\omega, \end{equation}
and
\begin{equation}\label{SUSDKin} \omega-kv_{Fu}=\Delta+\delta\omega, \end{equation}
where $\Delta=k(v_{Fd}-v_{Fu})$.

In this regime the dispersion appears as follows
$$\omega =kv_{Fd}\Biggl\{1+ 2\exp\biggl[-2-\frac{2}{3}\frac{k^{2}v_{Fd}^{2}}{\omega_{Ld}^{2}}$$
\begin{equation}\label{SUSDKin DD Zero s with up}  -\frac{n_{0u}^{\frac{1}{3}}}{n_{0d}^{\frac{1}{3}}}\biggl(2 +\ln\frac{n_{0d}^{\frac{1}{3}}-n_{0u}^{\frac{1}{3}}}{n_{0d}^{\frac{1}{3}}+n_{0u}^{\frac{1}{3}}}\biggr)\biggr]\Biggr\}. \end{equation}

\section{Conclusions}

Method of separate spin evolution quantum kinetics, which separately describes spin-up and spin-down electrons, has been developed. This method has been applied to rederivation of spectrum of the  Langmuir waves and the SEAWs obtained earlier in terms of SSE-QHD. Regime of wave propagation parallel to the external magnetic field has been considered at calculations of spectrum of magnetised spin-1/2 quantum plasmas. Contribution of ions dynamics in dispersion of SEAW has been considered. Calculation of the Landau damping of the SEAW has been performed. Influence of separated spin evolution on real and imaginary parts of spectrums of ion-acoustic waves and zeroth sound have been found.

Calculation of the Landau damping of the SEAWs has demonstrated that the SEAWs are weakly damped waves. Thus we have shown that hydrodynamic calculations of the real part of spectrum of the SEAWs were reasonable.

We have presented fundamental applications of the separates spin evolution quantum kinetics method. Furthermore, this method, along with the developed earlier SSE-QHD, creates strong background for research of spin-1/2 quantum plasmas. It open possibilities for more detailed analysis of various effects in quantum plasmas then usual spin-1/2 QHD or similar quantum kinetics.

%%%%%%%%%%%%%%%%%%%%%%%%%%%%%%%%%%%%%%%%%%%%%

\begin{acknowledgements}
The author thanks Professor L. S. Kuz'menkov for fruitful discussions.
\end{acknowledgements}

%\section{Appendix}

\end{document}